\documentclass[10pt]{article}
\usepackage{amsmath}
\usepackage{graphicx}
\usepackage{amsfonts}
\usepackage{amssymb}
\usepackage{epsf}
\usepackage{latexsym}

\def\80{\hspace{0.8in}}

\def\bq{\bf q}
\def\bR{\bf R}
\def\br{\bf r}
\def\brho{\mbox{\boldmath$\rho$}}
\newcommand{\be}{\begin{enumerate}}
\newcommand{\ee}{\end{enumerate}}
\newcommand{\bi}{\begin{itemize}}
\newcommand{\ei}{\end{itemize}}
\newcommand{\bd}{\begin{description}}
\newcommand{\ed}{\end{description}}
\def\beq{\begin{equation}}
\def\eeq{\end{equation}}
\def\bea{\begin{eqnarray}}
\def\eea{\end{eqnarray}}
\def\hat{\widehat}

\def\pa{\partial}
\def\d{\textrm{d}}

\def\ttH{\mbox{\tt H}}

\def\ttD{\mbox{\tt D}}

\def\cr{\mbox{\scriptsize{\bf $\mbox{ } \times \mbox{ }$}}}

\def\mh{\mbox{h}}

\def\mn{\mbox{n}}

\def\mz{\mbox{z}}

\def\mC{\mbox{C}}
\def\mD{\mbox{D}}
\def\mE{\mbox{E}}

\def\mM{\mbox{M}}

\def\mZ{\mbox{Z}} 
\def\sa{\mbox{\scriptsize a}}
\def\sb{\mbox{\scriptsize b}}
\def\sc{\mbox{\scriptsize c}}

\def\se{\mbox{\scriptsize e}}

\def\sh{\mbox{\scriptsize h}} 

\def\sj{\mbox{\scriptsize j}} 

\def\sll{\mbox{\scriptsize l}}  
\def\sm{\mbox{\scriptsize m}}
\def\sn{\mbox{\scriptsize n}} 
\def\so{\mbox{\scriptsize o}} 
\def\sp{\mbox{\scriptsize p}}

\def\sss{\mbox{\scriptsize s}}
\def\su{\mbox{\scriptsize u}}

\def\sC{\mbox{\scriptsize C}}
\def\sD{\mbox{\scriptsize D}}
\def\sE{\mbox{\scriptsize E}}
\def\sF{\mbox{\scriptsize F}}

\def\sH{\mbox{\scriptsize H}}

\def\sK{\mbox{\scriptsize K}}

\def\sN{\mbox{\scriptsize N}}

\def\sR{\mbox{\scriptsize R}}
\def\sS{\mbox{\scriptsize S}}
\def\sT{\mbox{\scriptsize T}}

\def\eph(B){\mbox{\scriptsize emergent(LMB)}}

\def\te{\mbox{\tiny e}}

\def\tm{\mbox{\tiny m}}

\def\tD{\mbox{\tiny D}}

\def\tH{\mbox{\tiny H}}

\def\tK{\mbox{\tiny K}}

\def\tT{\mbox{\tiny T}}

\def\fE{\mbox{\sffamily E}}

\def\fI{\mbox{\sffamily I}}

\def\fP{\mbox{\sffamily P}}

\def\fR{\mbox{\sffamily R}}
\def\fS{\mbox{\sffamily S}}
\def\fT{\mbox{\sffamily T}}

\def\fV{\mbox{\sffamily V}}
\def\fW{\mbox{\sffamily W}}

\renewcommand{\H}{{\cal H}}                 % !!! Command redefined !!!

\def\bn{\mbox{\bf n}}

\def\bP{\mbox{\bf P}}

\begin{document}

\begin{titlepage}

\begin{center}

{\Large\bf RELATIONAL QUADRILATERALLAND INTERPRETATION}

\vspace{.1in}

{\LARGE\bf OF $\mathbb{CP}^2$ AND QUOTIENTS}

\vspace{.2in}

{\bf Edward Anderson}$^1$ 

\vspace{.2in}

\large {\em $^1$ Astroparticule et Cosmologie, Universit\'{e} Paris 7 Diderot} \normalsize

\vspace{.2in}

\end{center}

\begin{abstract}

I investigate qualitatively significant regions of the configuration space for the classical and quantum mechanics of the relational 
quadrilateral in 2-d.  
This is relational in the sense that only relative ratios of separations, relative angles and relative times are significant.
Such relational particle mechanics models have many analogies with the geometrodynamical formulation of general relativity.  
Thus, they are suitable as toy models for  studying 
1) problem of time in quantum gravity strategies, 
in particular timeless, semiclassical and histories theory approaches and combinations of these. 
2) Various other quantum-cosmological issues, such as structure 
formation/inhomogeneity and the significance of uniform states.    
The relational quadrilateral is more useful in these respects than previously investigated simpler 
RPM's as it possesses linear constraints, nontrivial subsystems and its configuration 
space is a nontrivial complex-projective space.  
In this paper, I investigate the submanifold of collinear configurations, the submanifolds with a single-particle collision, 
the square configurations and regions of the configuration space for which such as approximate collinearity,  
approximate squareness and approximate triangularity hold.   
I consider both mirror image identified and unidentified shapes, as well as both distinguishable and indistinguishable 
particle labels.    
I find the following $\mathbb{CP}^2$ coordinate systems to be useful for these purposes. 
1) Kuiper's coordinates, which, for quadrilateralland, are magnitudes of the Jacobi vectors and anisoscelesnesses of the 
triangles formed between them. 
2) Gibbons--Pope type coordinates, which, for quadrilateralland, are the sum and difference of relative
angles between subsystems, the ratio of size of 2 subsystems and the proportion of universe model occupied by the subsystems.  

\end{abstract}

\vspace{0.2in} 

\noindent Seminar II on relational quadrilaterals


\vspace{0.2in}  

$^1$  edward.anderson@apc.univ-paris7.fr (Work started at DAMTP Cambridge and continued at UAM Madrid)

% what is significance of democratic invariants in QIT?  

\end{titlepage}

%==========================================================================================================
\section{Introduction}
%==========================================================================================================

Relational particle mechanics (RPM) are mechanics in which only relative times, relative angles and 
(ratios of) relative separations are physically meaningful.  
Scaled RPM was set up in \cite{BB82} by Barbour and Bertotti and pure-shape (i.e. scalefree, and so involving 
just ratios) RPM was set up in  and in \cite{B03} by Barbour.   
These theories are relational in Barbour's sense as opposed to Rovelli's \cite{B94I, EOT, Rovellibook, 
08I, FileR}, via obeying the following postulates. 

\mbox{ }

\noindent 1) RPM's are {\it temporally relational} \cite{BB82, RWR, FORD}.  
I.e. they do not possess any meaningful primary notion of time for the whole (model) universe. 
This is implemented by using actions that are manifestly reparametrization invariant while also being 
free of extraneous time-related variables [such as Newtonian time or the lapse in General Relativity 
(GR)].   
These actions are of the form
\beq
\fI = 2\int\d s \sqrt{\fE - \fV } \mbox{ } . 
\label{Actio}
\eeq
(Jacobi-type actions \cite{Lanczos} for mechanics, which are closely related to 
Baierlein--Sharp--Wheeler-type actions \cite{BSW, RWR} for GR in geometrodynamical form.) 
This reparametrization invariance then directly produces primary constraints quadratic in the momenta.  
In the GR case, the constraint arising thus is the super-Hamiltonian constraint,\footnote{The notation 
%%%%%%%%%%%%%%%%%%%%%%%%%%%%%%%%%%%%%%%%%%%%%%%%%%%%%%%%%%%%%%%%%%%%%%%%%%%%%%%%%%%%%%%%%%%%%%%%%%%%%%%%%%%%%%%%%%%%%%%%%%%%%%%  
for this paper is as follows.
I take $a$, $b$, $c$ as particle label indices running from 1 to $N$ for particle positions,  
$e$, $f$, $g$ as particle label indices running from 1 to $n = N - 1$ for relative position variables,  
$I$, $J$, $K$ as indices running from 1 to $n - 1$ and 
$i$, $j$, $k$ as spatial indices.  
I term 1-$d$ RPM's $N${\it -stop metrolands} from their configurations looking like public transport line maps, 
and 2-$d$ ones $N${\it -a-gonlands}, the first nontrivial two of which are {\it triangleland} and 
{\it quadrilateralland}. 
$p$, $q$, $r$ are then relational space indices (running from 1 to 3 for triangleland and from 1 to 5 for quadrilateralland), and 
$u$, $v$, $w$ as shape space indices (running from 1 to 4 for quadrilateralland).
$h_{ij}$ is the spatial 3-metric, with determinant $h$, undensitized supermetric $N_{ijkl} = h_{ik}h_{jl} - 
\frac{1}{2}h_{ij}h_{kl}$, conjugate momentum $\pi^{ij}$, covariant derivative $D_{i}$ and Ricci scalar $R$. 
$\fE$ is the total energy, $\fV$ is the potential energy, and ${\bR}^e$ are relative Jacobi interparticle (cluster) 
coordinate vectors  (see Figure 1 in Sec 2 for further specification of what these are), 
with associated masses $\mu_e$ and conjugate momenta ${\bP}_e$.}
%%%%%%%%%%%%%%%%%%%%%%%%%%%%%%%%%%%%%%%%%%%%%%%%%%%%%%%%%%%%%%%%%%%%%%%%%%%%%%%%%%%%%%%%%%%%%%%%%%%%%%%%%%%%%%%%%%%%%%%%%%%%%%%  
\beq
{\cal H} := N_{ijkl}\pi^{ij}\pi^{kl}/\sqrt{h} - \sqrt{h}\{R - 2\Lambda\} = 0 \mbox{ } , 
\eeq 
whilst for RPM's it gives an energy constraint, 
\beq
\ttH := \sum\mbox{}_{\mbox{}_{\mbox{\scriptsize $e$ = 1}}}^n  P_e\mbox{}^2/2\mu_e + \fV = \fE \mbox{ } .    
\eeq
2) RPM's are {\it configurationally relational}.  
This can be thought of as some group $G$ of transformations that act on the theory's configuration space $Q$ 
being taken to be physically meaningless \cite{BB82, RWR, FORD, FEPI, Cones}.   
Configurational relationalism can be approached by \cite{BB82, B03} indirect means that amount to 
working on the principal bundle P($Q$, $G$).  
In this setting, linear constraints arise from varying with respect to the $G$-generators.  
In the case of GR, such a variation with respect to the generators of the 3-diffeomorphisms 
gives the super-momentum constraint 
\beq
{\cal L}_i := -2D_j{\pi^j}_i = 0 \mbox{ } ,
\eeq 
whilst for RPM's variation with respect to the generators of the rotations gives that the 
total angular momentum for the model universe as a whole is constrained to be zero, 
\beq
{\bf L} := \sum\mbox{}_{\mbox{}_{\mbox{\scriptsize $e$ = 1}}}^n {\bR}^e\cr {\bP}_e =0 \mbox{ } . 
\label{ZAM}
\eeq 
In the case of pure-shape RPM, variation with respect to the generator of the dilations gives 
that the total dilational momentum of the model universe is constrained to be zero, 
\beq 
\ttD := \sum\mbox{}_{\mbox{}_{\mbox{\scriptsize $e$ = 1}}}^n   {\bR}^e\cdot {\bP}_e = 0 \mbox{ } . 
\eeq 
Next, elimination of the $G$-generator variables from the Lagrangian form of these constraints sends one 
to the desired quotient space $Q/G$.
In 1- and 2-$d$, these RPM's can be obtained alternatively by working directly on $Q/G$ \cite{FORD, Cones} 
(see Sec 3 to 5 for further details). 
This involves applying a `Jacobi--Synge' construction of the natural mechanics associated with a geometry, 
to Kendall's shape spaces \cite{Kendall} for pure-shape RPM's or to the cones over these \cite{Cones} for 
the scaled RPM's.   
 
\mbox{ }    

RPM's are further motivated as useful toy models \cite{K92, EOT, RWR, Kieferbook, 
06I06IISemiclIGrybBanal, ScaleQM, MGM, Records, FEPI, AF, Cones} of yet further features of GR cast in 
geometrodynamical form.  
RPM's are similar in complexity and in number of parallels with GR to minisuperspace \cite{Mini} and to  
2 + 1 GR \cite{Carlip} are, though each such model differs in how it resembles GR, so that such models  
offer complementary perspectives and insights.\footnote{E.g. \cite{K92, I93, Kieferbook} use a wide  
%%%%%%%%%%%%%%%%%%%%%%%%%%%%%%%%%%%%%%%%%%%%%%%%%%%%%%%%%%%%%%%%%%%%%%%%%%%%%%%%%%%%%%%%%%%%%%%%%%%%%%%%
variety of such models in discussing the Problem of Time in Quantum Gravity and other foundational issues.}
%%%%%%%%%%%%%%%%%%%%%%%%%%%%%%%%%%%%%%%%%%%%%%%%%%%%%%%%%%%%%%%%%%%%%%%%%%%%%%%%%%%%%%%%%%%%%%%%%%%%%%%%
%
In particular 

\noindent I) The analogies with GR continue at the level of configuration spaces (see Fig \ref{Fig2} in Sec 2).

\noindent II) RPM's energy constraint parallels the GR super-Hamiltonian constraint ${\cal H}$ in leading to 
the frozen formalism facet of the Problem of Time.   
This notorious problem occurs because `time' takes a different meaning in each of GR and ordinary 
quantum theory.  
This incompatibility underscores a number of problems with trying to replace these two branches with a 
single framework in situations in which the premises of both apply, such as in black holes or the very 
early universe.  
One facet of the Problem of Time that then appears in attempting canonical quantization of GR is due to ${\cal H}$ 
being quadratic but not linear in the momenta, which feature and consequence are shared by $\ttH$.
Then elevating ${\cal H}$ to a quantum equation produces a stationary i.e timeless or frozen wave 
equation: the  Wheeler-DeWitt equation 
\beq
\hat{\cal H}\Psi = 0 \mbox{ } , 
\eeq 
instead of ordinary QM's time-dependent one, 
\beq
i\hbar\pa\Psi/\pa t = \hat{H}\Psi
\eeq 
(where I use $\Psi$ for the wavefunction of the universe, $H$ to denote a Hamiltonian and $t$ for 
absolute Newtonian time).  
See \cite{K92, I93, APOT} for other facets of the Problem of Time.  

\noindent III) The nontrivial linear constraints parallel the GR momentum constraint (and are absent 
for minisuperspace), which is the cause of a number of further difficulties in various approaches to the 
Problem of Time.    

\noindent By parallels I), II) and III), RPM큦 are appropriate as toy models for a large number of Problem 
of Time approaches [see the Conclusion for more details].  
Other useful applications of RPM's not covered by minisuperspace models include

\noindent IV) that RPM's are useful for the qualitative study of the quantum-cosmological origin of 
structure formation/ inhomogeneity. 
(Scaled RPM's are a tightly analogous, simpler version of Halliwell and   Hawking's \cite{HallHaw} model 
for this; moreover scalefree RPM's such as this paper's occurs as a subproblem within scaled RPM's, 
corresponding to the light fast modes/inhomogeneities.) 

\noindent V) RPM's are likewise useful for the study of correlations between localized subsystems of a given instant. 

\noindent VI) RPM's also allow for a qualitative study of notions of uniformity/of maximally or highly uniform states 
in classical and quantum Cosmology, which are held to be conceptually important notions in these subjects.  

\mbox{ }

Some of the analogies with GR already hold for the 1-d $N$-stop metroland RPM's, which have the particularly tractable 
\{$N$ -- 2\}-spheres $\mathbb{S}^{N - 2}$ as shape spaces.
2-$d$ suffices for almost all the analogies with GR to hold whilst still keeping the mathematics manageable; the 
shape spaces are the complex projective spaces $\mathbb{CP}^{N - 2}$.    
$N$-stop metroland and triangleland RPM's have already been covered in \cite{AF, ScaleQM, 08I, 08II, +tri, 08III}. 
Thus, we are now looking at the next $N$-a-gonland: the $N$ = 4 case, {\it quadrilateralland}.  
This is valuable through its simultaneously possessing the following features. 

\noindent 
1) Nontrivial linear constraints.   

\noindent 
2) Nontrivial clustering/inhomogeneity/structure (i.e. midisuperspace-like features rendering 
it suitable as a qualitative toy model of Halliwell and Hawking's quantum cosmological origin of 
structure formation in the universe). 

\noindent 
3) Relationally nontrivial non-overlapping subsystems and hierarchies of nontrivial subsystems.  
[These are useful features for timeless approaches as well as for a less trivial structure formation 
than in the triangleland case.]

\noindent 
Also, quadrilateralland possesses nontrivial complex projective space mathematics (triangleland atypically 
simplifies via $\mathbb{CP}^1 = \mathbb{S}^2$, whereas quadrilateralland 
is much more mathematically typical for an $N$-a-gonland.)

I emphasize that the study of quadrilateralland is still at an early stage.      
In a previous paper \cite{QShape}, I considered i) shape quantities for quadrilateralland that parallel the 
Dragt-type coordinates \cite{Dragt} that were so useful for triangleland \cite{+tri, 08III} (and are closely 
related to the Hopf map). 
ii) The 
clustering-independent (`democratically invariant' \cite{ZickACG86LR, LR95}) properties for this. 
iii) This leads to a quantifier of uniformity for model-universe configurations.  
The present paper is the first instance in RPM literature involving detailed treatment of indistinguishable 
particles as well as the first paper on  submanifolds within $\mathbb{CP}^2$ interpreted in quadrilateralland terms.  
These two papers, and a third concerning the interpretation of $\mathbb{CP}^2$'s $SU(3)$ of conserved quantities 
in quadrilateralland terms \cite{QCons}, are important prerequisites for 
the subsequent study of the classical and quantum mechanics of quadrilateralland \cite{Quad1}. 
These are then useful as a nontrivial model of the Problem of Time in Quantum Gravity and of various 
other foundational and qualitative issues in Quantum Cosmology (extending e.g. \cite{08III, SemiclIII, FileR}).

A distinct physical interpretation of $\mathbb{CP}^2$ is as qtrits in Quantum Information Theory.  
qtrits and q$n$its in general are motivated by there being much more information storage in these than in qbits.
$\mathbb{CP} ^{N - 2}$ occurs in this application as the space of quantum states.
Thus additionally qtrits can also be interpreted in terms of quadrilaterals and qu$n$its in terms of $N$-a-gons for $N = n + 1$.

For more uses of $\mathbb{CP}^2$, see \cite{MacFarlane, QCons}.

\mbox{ } 

In Sec 2 I provide coordinate systems and types of configuration space that are useful in the study of 
the general relational particle mechanics models.  
I then consider details of the configuration spaces for 4-stop metroland (Sec 3) and triangleland (Sec 4),  
alongside further coordinatizations useful in these specific cases.
These (and extensions of them) then feature within the paper's main focus, quadrilateralland (Sec 5).
For $\mathbb{CP}^2$, a useful set of redundant coordinates are Kuiper coordinates \cite{Kuiper}.  
These are interpretable in the quadrilateralland setting as magnitudes of the Jacobi vectors alongside the 
inner products between pairs of these. 
The latter amount to relative angles, interpretable as the departures from 
isoscelesness of the coarse-graining triangles associated with the quadrilateral (or departure from rhombicness 
of the coarse-graining parallelogram).    
I use these to determine the quadrilateralland counterpart of the split of triangleland's sphere into two 
hemispheres of mirror-image configurations that join along an equator of collinear configurations 
(which is one of \cite{+tri}'s most important and useful results).  
Another useful set of coordinates  for $\mathbb{CP}^2$ 
are Gibbons--Pope type coordinates \cite{GiPo} (intrinsic to $\mathbb{CP}^2$ itself).  
In this paper I interpret these in quadrilateralland terms, showing how spherical coordinates for both 
4-stop metroland and triangleland extend neatly into this this set of quadrilateralland coordinates.  
These give a clear way of seeing how the $\mathbb{CP}^2$ metric reduces to a $\mathbb{S}^2$ one when two of 
the particles collide so that one is left with a triangular configuration. 
I also investigate the square configurations and a weaker notion of highly uniform states using both of the 
above coordinate systems.   
In the conclusion (Sec 6), I include a further application of the Gibbons--Pope type coordinates to quadrilateralland:  
for computing integrals over geometrically/physically significant regions of the quadrilateralland shape space 
(these are relevant to various approaches to the Problem of Time) and comment on the $N$-a-gonland extension of 
the present paper.

%========================================================================================================
\section{Coordinate systems used in this paper}  
%========================================================================================================

%FFFFFFFFFFFFFFFFFFFFFFFFFFFFFFFFFFFFFFFFFFFFFFFFFFFFFFFFFFFFFFFFFFFFFFFFFFFFFFFFFFFFFFFFFFFFFFFFFFFFFFF
{           \begin{figure}[ht]
\centering
\includegraphics[width=0.7\textwidth]{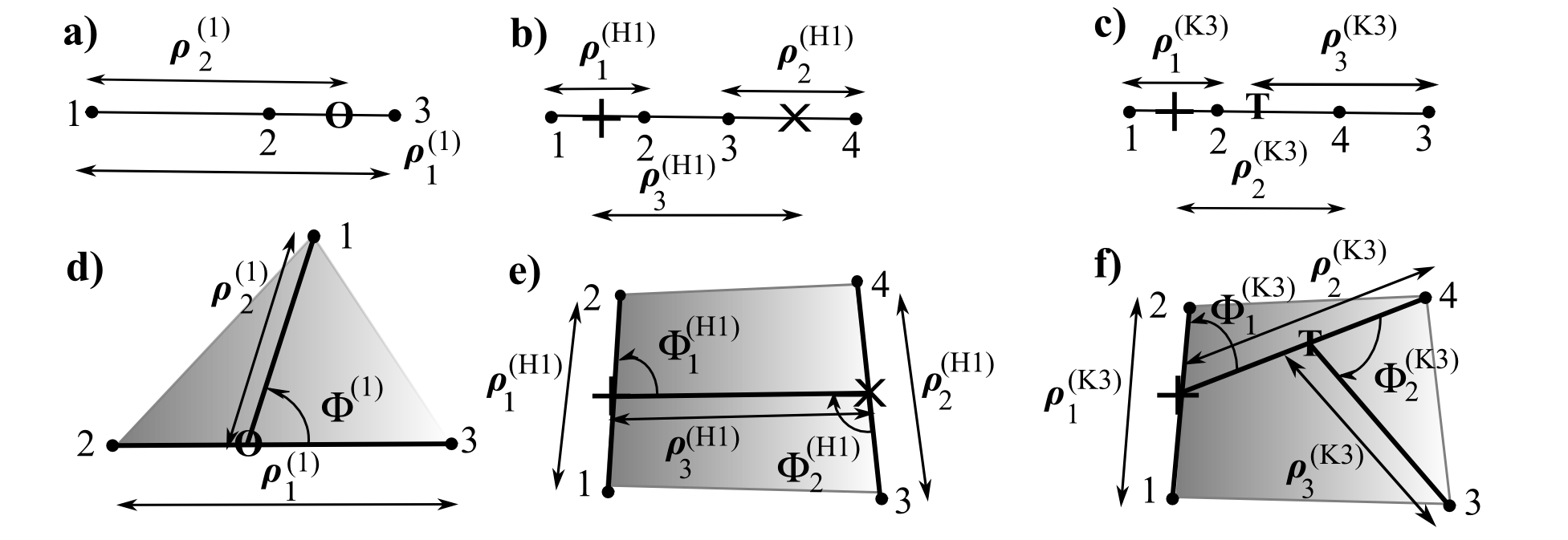}
\caption[Text der im Bilderverzeichnis auftaucht]{  \footnotesize{O, +, $\times$ and T denote COM(23), 
COM(12), COM(34) and COM(124) respectively, where COM(ab) is the centre of mass of particles a and b.
a) For 3 particles in 1-$d$, one particular choice of mass-weighted relative Jacobi coordinates are 
as indicated. 
\noindent I use (Hb) as shorthand for \{ab, cd\} i.e. the clustering (partition into subclusters) into two pairs 
\{ab\} and \{cd\}, and (Ka) as shorthand for 
\{\{cd, b\}, a\} i.e. the clustering into a single particle a 
and a triple \{cd, b\} which is itself partitioned into a pair cd and a single particle b.
In each case, a, b, c, d form a cycle.
In this paper, I take clockwise and anticlockwise labelled triangles to be distinct, 
i.e. I make the plain rather than mirror-image-identified choice of set of shapes.
I also assume equal masses for simplicity.  
For 4 particles in 1-$d$, b) and c) give, respectively, a particular choice of mass-weighted relative 
Jacobi H-coordinates [the squashed version of the obvious H-shape in e)] and of 
Jacobi K-coordinates [the squashed version of the obvious K-shape in f)].
This paper also represents a fair amount of RPM work done for the first time in Jacobi K-coordinates  
Later references to H and K coordinates refer explicitly to the (H1) and (K3) cases depicted above; I 
drop these labels to simplify the notation.  
d) For 3 particles in 2-$d$, one particular choice of mass-weighted relative Jacobi coordinates are 
as indicated. 
I furthermore define $\Phi^{(\sa)}$ as the `Swiss army knife' angle 
$\mbox{arccos}\big( \brho_1^{(\sa)} \cdot \brho_3^{(\sa)} / \rho_1^{(\sa)} \rho_3^{(\sa)} \big)$.
For 4 particles in 2-$d$ e) and f) give, respectively, a particular choice of mass-weighted relative Jacobi 
H-coordinates [with $\Phi_1^{(\sH\sb)}$ and $\Phi_2^{(\sH\sb)}$ as the `Swiss army knife' angles 
$\mbox{arccos}\big(\brho_1^{(\sH\sb)}\cdot\brho_3^{(\sH\sb)}/\rho_1^{(\sH\sb)}\rho_3^{(\sH\sb)}\big)$ 
and 
$\mbox{arccos}\big(\brho_2^{(\sH\sb)}\cdot\brho_3^{(\sH\sb)}/\rho_2^{(\sH\sb)}\rho_3^{(\sH\sb)}\big)$ 
respectively] and of 
K-coordinates [with $\Phi_1^{(\sK\sa)}$ and $\Phi_2^{(\sK\sa)}$ as the `Swiss army knife' angles  
$\mbox{arccos}\big( \brho_1^{(\sK\sa)}\cdot\brho_2^{(\sK\sa)}/\rho_1^{(\sK\sa)}\rho_2^{(\sK\sa)} \big) $ 
and  
arccos$\big( \brho_2^{(\sK\sa)}\cdot\brho_3^{(\sK\sa)}/\rho_2^{(\sK\sa)}\rho_3^{(\sK\sa)} \big)$   respectively].     }        } 
\label{Fig1} \end{figure}         } 
%FFFFFFFFFFFFFFFFFFFFFFFFFFFFFFFFFFFFFFFFFFFFFFFFFFFFFFFFFFFFFFFFFFFFFFFFFFFFFFFFFFFFFFFFFFFFFFFFFFFFFFFF

\noindent $N$ particles in dimension $d$ has Cartesian configuration space $\mathbb{R}^{dN}$.
Rendering absolute position irrelevant (e.g. by passing from particle position coordinates 
${\bq}^a$ to any sort of relative coordinates) leaves one on the configuration space {\it relative 
space}, $\fR = \mathbb{R}^{2n}$, for $n$ = $N$ -- 1.
The most convenient sort of coordinates for this are {\it relative Jacobi coordinates} \cite{Marchal} 
${\bR}^e$.  
These are combinations of relative position vectors ${\br}^{ab} = {\bq}^b - {\bq}^a$ between particles 
into inter-particle cluster vectors such that the kinetic term is diagonal.  
Relative Jacobi coordinates have associated particle cluster masses $\mu_e$. 
In fact, it is tidier to use {\it mass-weighted} relative Jacobi coordinates $\brho^e = \sqrt{\mu_e}
{\bR}^e$ (Fig \ref{Fig1}). 
The squares of the magnitudes of these are the partial moments of inertia $I^e = \mu_e|{\bR}^e|^2$.  
I also denote $|\brho^e|$ by $\rho^e$, $I^e/I$ by $N^e$ for $I$ the moment of inertia of the system, and
$\brho^e/\rho$ by $\bn^e$ for $\rho = \sqrt{I}$ the {\it hyperradius}.   
If one quotients out the rotations also, one is on {\it relational space} ${\cal R}(N, d)$.    
If one furthermore quotients out the scalings, one is on {\it shape space} $\fS(N, d)$. 
If one quotients out the scalings but not the rotations, one is on {\it preshape space} \cite{Kendall} $\fP(N,d)$  
[see Fig 2a)].

%FFFFFFFFFFFFFFFFFFFFFFFFFFFFFFFFFFFFFFFFFFFFFFFFFFFFFFFFFFFFFFFFFFFFFFFFFFFFFFFFFFFFFFFFFFFFFFFFFFFFFFF
{            \begin{figure}[ht]
\centering
\includegraphics[width=0.75\textwidth]{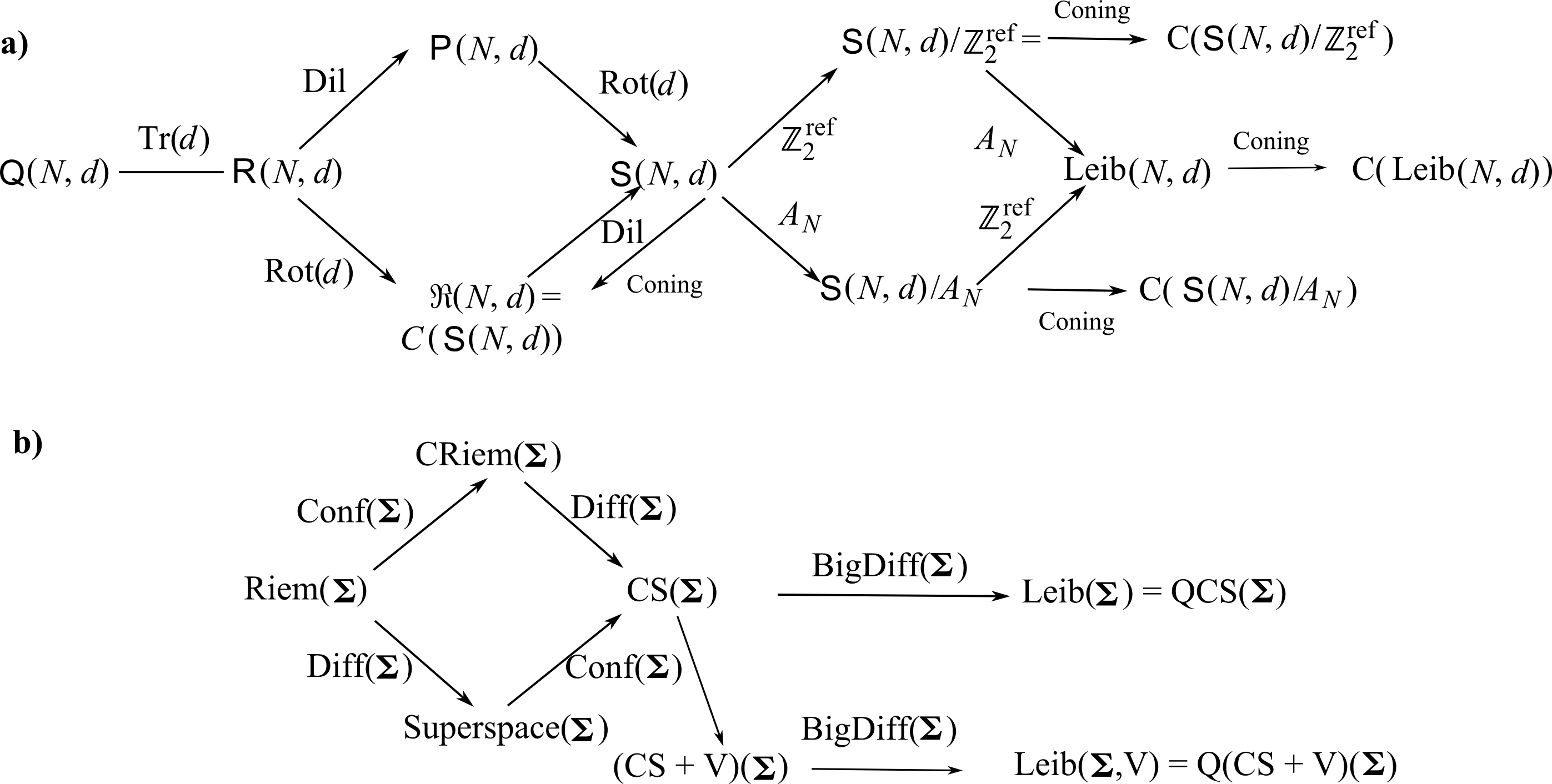}
\caption[Text der im Bilderverzeichnis auftaucht]{        \footnotesize{a) the sequence of 
configuration spaces of this paper. 
b) The corresponding sequence of configuration spaces for GR.       
Riem($\Sigma$) is the space of positive-definite 3 metrics on a fixed topology $\Sigma$, taken to be a compact without boundary 
one for simplicity.  
These most naturally correspond to relative space $\fR$($N, d$).  
Diff($\Sigma$) and Conf($\Sigma$) are the corresponding 3-diffeomorphisms and conformal transformations; 
these most naturally correspond to the rotations Rot($d$) and dilations Dil respectively.  
Superspace($\Sigma$) is Wheeler's notion \cite{Battelle} (entirely unrelated to the later use of this word 
in the context of supersymmetry). 
CRiem($\Sigma$) is pointwise superspace and  CS($\Sigma$) is conformal superspace.
Superspace($\Sigma$), CRiem($\Sigma$) and CS($\Sigma$) most naturally correspond to relational space 
${\cal R}(N, d)$, preshape space P($N, d$) and shape space S($N, d$) respectively.  
(RPM's also admit analogies with conformal/York-type \cite{York73York74ABFOABFKO} initial value problem 
formulations of GR, with the conformal 3-geometries playing here an analogous role to the pure shapes.)
(CS + V)($\Sigma$) is conformal superspace to which has been adjoined a single global degree of freedom: the spatial volume 
of the universe. 
CS($\Sigma$) and (CS + V)($\Sigma$) have on a number of occasions been claimed to 
be the space of true dynamical degrees of freedom of GR \cite{York73York74ABFOABFKO, NewBO}.
The quotienting out of large diffeomorphisms gives the notion of quantum superspace and quantum conformal superspace 
[QCS($\Sigma$)] as per \cite{FM96}.  
This corresponds most naturally to identifying mirror image shapes and enforcing particle indistinguishability.
Leib($N, d) = \fS(N, d)/S_N$ is then the analogue of QCS($\Sigma$) and C(Leib($N,d$)) is then in some ways the analogue of 
quantum CS + V.  
It is named thus as the most very Leibnizian of the possible configuration spaces for mechanics with equal particle masses. } }
\label{Fig2} \end{figure}          }
%FFFFFFFFFFFFFFFFFFFFFFFFFFFFFFFFFFFFFFFFFFFFFFFFFFFFFFFFFFFFFFFFFFFFFFFFFFFFFFFFFFFFFFFFFFFFFFFFFFFFFFFF

For $N$ particles in  1-$d$, preshape space is $\mathbb{S}^{N - 2}$ (or some piece thereof, see Sec 3) 
which coincides with shape space. 
For $N$ particles in 2-$d$, preshape space is $\mathbb{S}^{nd - 1}$ and shape space is 
$\mathbb{CP}^{N - 2}$ (provided that the plain choice of set of shapes is made).  
The 3-particle case of this is, moreover, special, by $\mathbb{CP}^1 = \mathbb{S}^2$.
Also, relational space ${\cal R}(N, d)$ is equal to \cite{Cones} {\it the cone} over shape space, denoted by $\mC(\fS(N, d))$. 
[At the topological level, for C(X) to be a cone over some topological manifold X, 
\beq
\mbox{C(X) = X $\times$ [0, $\infty$)/\mbox{ }$\widetilde{\mbox{ }}$} \mbox{ } , 
\eeq
where $\widetilde{\mbox{ }}$ means that all points of the form \{p $\in$ X, 0 $\in [0, \infty)$\} are 
`squashed' or identified to a single point termed the {\it cone point}, and denoted by 0. 
For what a cone further signifies at the level of Riemannian geometry, see below.]  
In particular, ${\cal R}(N, 1) = \mC(\mathbb{S}^{N - 2} )= \mathbb{R}^{n}$ and 
${\cal R}(N, 2)$ = C$(\mathbb{CP}^{N - 2})$ [among which additionally $\mC(\mathbb{CP}^1) 
= \mC(\mathbb{S}^2) = \mathbb{R}^3$ at the topological level].  
Thus, this paper's quadrilateralland case is the first case with nontrivial 
complex-projective mathematics.

For 1-$d$, one has the usual \{$N$ -- 2\}-sphere metric on shape space, $\d^2s_{\sss\sp\sh\se}$, with 
then the Euclidean metric on the corresponding relational space, $\d^2s_{\sE\su\sc\sll} = \d\rho^2 + \rho^2\d^2s_{\sss\sp\sh\se}$.   
For 2-$d$, the kinetic metric on shape space is the natural Fubini--Study metric \cite{Kendall, FORD}, 
\beq
\d s_{\sF\sS}^2 = \big\{\{1 + |{\mZ}|_{\sc}^2\}|\d {\mZ}|_{\sc}^2 - |({\mZ} ,\d\overline{{\mZ}}\}_{\sc}|^2\big\}/
\{1 + |{\mZ}|_{\sc}^2\}^2  \mbox{ } .
\label{FS} 
\eeq
I explain the coordinates in use here as follows. 
Firstly, \{${\mz}^e$\} are, mathematically, complex 
homogeneous coordinates for $\mathbb{CP}^2$; 
I denote their polar form by $\mz^e = \rho^e\mbox{exp}({i\phi^{e}})$. 
Physically, these contain 2 redundancies and their moduli are the magnitudes of the Jacobi vectors whilst their 
arguments are angles between the Jacobi vectors and an absolute axis.  
Next, \{${\mZ}^I$\} are, mathematically, complex inhomogeneous coordinates for $\mathbb{CP}^2$; 
I denote their polar form by $\mZ^{I} = {\cal R}^{I}\mbox{exp}({i\Phi^{I}})$.  
These are independent ratios of the $\mz^{e}$, and so, physically their magnitudes ${\cal R}^{I}$ are ratios of 
magnitudes of Jacobi vectors, and their arguments $\Phi^I$ are now angles between Jacobi vectors, which are 
entirely relational quantities.   
Also, I use $|{\mZ}|_{\sc}^2 := \sum_{I}|{\mZ}^{I}|^2$, $( \mbox{ } , \mbox{ } )_{\sc}$ for the corresponding inner 
product, overline to denote complex conjugate and $|\mbox{ }|$ to denote complex modulus.
Note that using the polar form for the $\mZ^I$, the line element and the corresponding kinetic term can be 
cast in a real form. 
The kinetic metric on relational space in scale-shape split coordinates is then of the cone form 
\beq
\d s_{\sC(\sF\sS)}^2 = \d \rho^2 + \rho^2\d s_{\sF\sS}^2 \mbox{ }  
\eeq
[for $\d s^2$ is the line element of X itself and $\rho$ a suitable `radial variable'\footnote{In the spherical 
%%%%%%%%%%%%%%%%%%%%%%%%%%%%%%%%%%%%%%%%%%%%%%%%%%%%%%%%%%%%%%%%%%%%%%%%%%%%%%%%%%%%%%%%%%%%%%%%%%%%%%%%%%%%%%%%%%%%%%%%%
presentation of the triangleland case, coordinate ranges dictate that the radial variable is, rather, $I$.  
Also note that, whilst this cone is topologically $\mathbb{R}^3$, the metric 
it comes equipped with is {\sl no}t the flat metric (though it is exploitably {\it conformally flat} \cite{08I, 08III}).}
%%%%%%%%%%%%%%%%%%%%%%%%%%%%%%%%%%%%%%%%%%%%%%%%%%%%%%%%%%%%%%%%%%%%%%%%%%%%%%%%%%%%%%%%%%%%%%%%%%%%%%%%%%%%%%%%%%%%%%%%%%
that parametrizes the [0, $\infty$), which is the distance from the cone point; 
such `cone metrics' are smooth everywhere except (possibly) at the troublesome cone point].

The action for RPM's in relational form is then (\ref{Actio}), which is a Jacobi-type action 
\cite{Lanczos} (thus complying with temporal relationalism), with, in 1- or 2-$d$,  
$\fT = \fT_{\sss\sp\sh\se}$ or $\fT_{\sE\su\sc\sll}$, $\fT = \fT_{\sF\sS}$ or $\fT_{\sC(\sF\sS)}$ 
built from the above selection of metrics (thus directly implementing configurational relationalism).

%=====================================================================================================================
\section{4-stop metroland}
%=====================================================================================================================

%FFFFFFFFFFFFFFFFFFFFFFFFFFFFFFFFFFFFFFFFFFFFFFFFFFFFFFFFFFFFFFFFFFFFFFFFFFFFFFFFFFFFFFFFFFFFFFFFFFFFFFF
{            \begin{figure}[ht]
\centering
\includegraphics[width=0.85\textwidth]{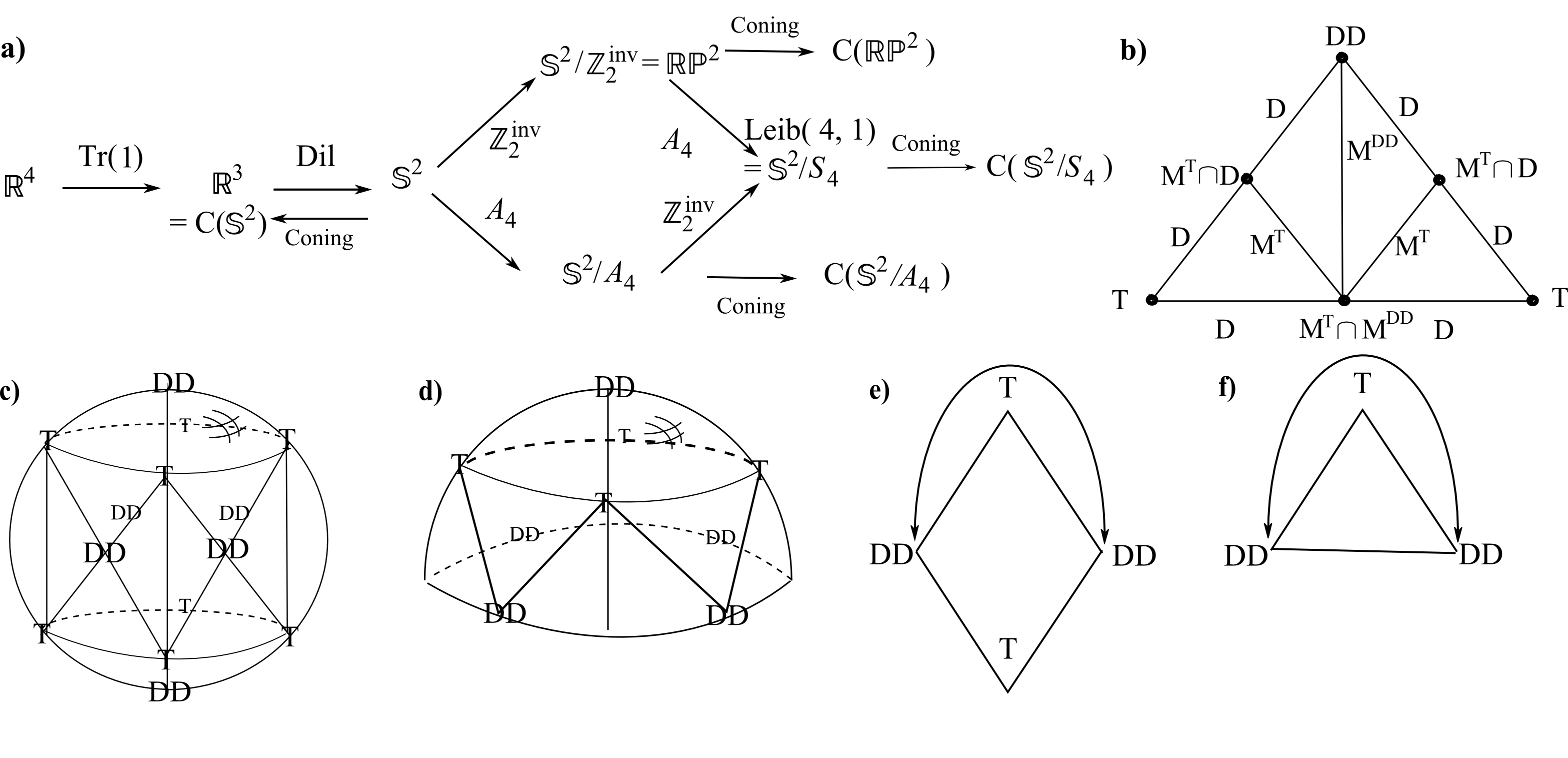}
\caption[Text der im Bilderverzeichnis auftaucht]{        \footnotesize{a) The sequence of configuration spaces for 4-stop metroland.
c) and d) are the tessellations for the distinguishable particles plain shape and mirror image identified cases respectively.  
b) illustrates the notions of merger within a single triangular face. 
There are $\mM^{\tT}$ arcs (for which the fourth 
particle is at the centre of mass of the other three) and $\mM^{\tD\tD}$ arcs (for which the centre of mass of 
2 pairs of particles coincide), as well as $\mM^{\tT} \bigcap \mD$ and $\mM^{\tD\tD} \bigcap \mM^{\tT}$ points.  
These are also the types of merger present in quadrilateralland.  
d) and e) are then the indistinguishable particle counterparts of these.}}
\label{Fig3} \end{figure}          }
%FFFFFFFFFFFFFFFFFFFFFFFFFFFFFFFFFFFFFFFFFFFFFFFFFFFFFFFFFFFFFFFFFFFFFFFFFFFFFFFFFFFFFFFFFFFFFFFFFFFFFFFF

The configuration space here for distinguishable particles and in the plain shape case is the sphere, decorated with 
the physical interpretation of Figs \ref{Fig3}c).
All the lines in Fig 3c) to f) are lines of double collisions, D.  
The DD points are double double collisions (two separate double collisions) whilst the T points are triple collisions.

At the level of configuration space geometry, the mirror-image-identification in space becomes inversion about the 
centre of the sphere. 
This gives rise to the real projective space, $\mathbb{RP}^2$ [Fig \ref{Fig3}c)].
Indistinguishability then involves quotienting out $S_4$ (permutations of the particles, which is isomorphic to 
the cube or octahaedron group acting on the DD's or the T's), as in Fig 3f).  
If there is no mirror-image identification, the quotienting out is rather by  $A_4$ (even permutations of the particles, 
isomorphic to the group of the cube or the octahaedron excluding one reflection operation), as in Fig 3e).

For later comparison with quadrilateralland [with matching enumeration, hence the absense of 2) and 3) below], 

\noindent 
1) A useful set of redundant coordinates for a surrounding flat Euclidean space (here equal to relational space) 
are, for 4-stop metroland, simply the three relative Jacobi coordinate magnitudes, $\mn^e$.  

\noindent 
4) Useful coordinates intrinsic to the shape space itself are the spherical polar coordinates $\theta$ and $\phi$, which 
have the following 4-stop metroland interpretations.
In Jacobi H-coordinates, 
\beq
\theta = \mbox{arctan}\big(\sqrt{\rho_1\mbox{}^2 + \rho_2\mbox{}^2}/\rho_3\big) \mbox{ } , \mbox{ } \mbox{ } 
\phi = \mbox{arctan}(\rho_2/\rho_1) \mbox{ } ,
\label{4StopPolars}
\eeq  
which are respectively a measure of the size of the universe's contents relative to the size of the 
whole model universe, and a measure of inhomogeneity among the contents of the universe (whether one of the 
constituent clusters is larger than the other one.)  
On the other hand, for Jacobi K-coordinates 
\beq
\theta = \mbox{arctan}\big(\sqrt{\rho_1\mbox{}^2 + \rho_2\mbox{}^2}/\rho_3\big) \mbox{ } , \mbox{ } \mbox{ } 
\phi = \mbox{arctan}(\rho_1/\rho_2) \mbox{ } ,
\label{4StopPolars2}
\eeq
which are respectively a measure the sizes of the \{12\} and \{T3\} clusters relative to the whole model universe 
and of the sizes of the \{12\} and \{T3\} clusters relative to each other.

%=========================================================================================================================
\section{Triangleland}
%=========================================================================================================================

%FFFFFFFFFFFFFFFFFFFFFFFFFFFFFFFFFFFFFFFFFFFFFFFFFFFFFFFFFFFFFFFFFFFFFFFFFFFFFFFFFFFFFFFFFFFFFFFFFFFFFFF
{            \begin{figure}[ht]
\centering
\includegraphics[width=0.77\textwidth]{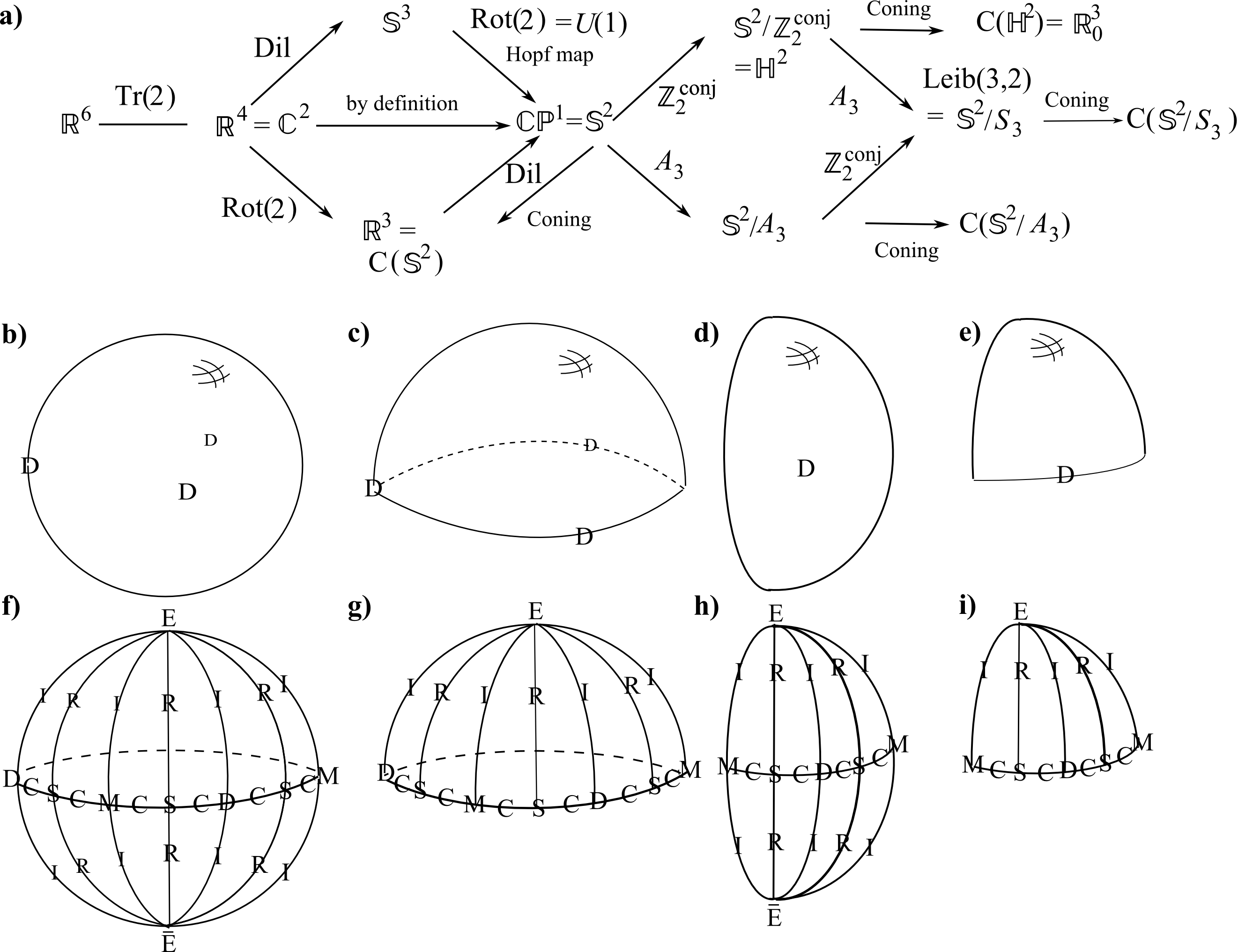}
\caption[Text der im Bilderverzeichnis auftaucht]{        \footnotesize{ a) The sequence of configuration spaces for triangleland.
$\mathbb{H}^2$ is the hemisphere with edge and $\mathbb{R}^3_0$ the half-space with edge.
b), c), d) and e) are the configuration spaces for, respectively, the distinguishable particle plain shape case, 
the distinguishable particle mirror-image-identified shape case,   the indistinguishable particle plain shape case and the indistinguishable particle mirror-image-identified shape case at the topological level. 
f), g), h) and i) are the tessellation for the distinguishable particle plain shape case, the distinguishable particle mirror-image-identified shape case, the indistinguishable particle plain shape case and the indistinguishable particle mirror-image-indentified 
shape case at the metric level.    }   }
\label{Fig4} \end{figure}          }
%FFFFFFFFFFFFFFFFFFFFFFFFFFFFFFFFFFFFFFFFFFFFFFFFFFFFFFFFFFFFFFFFFFFFFFFFFFFFFFFFFFFFFFFFFFFFFFFFFFFFFFFF

The configuration space here for distinguishable particles and in the plain shape case is the sphere, decorated 
as in Figs \ref{Fig4}b) and \ref{Fig4}f).
The labelled points and edges have the following geometrical/mechanical interpretations.  
E and $\bar{\mE}$ are the two mirror images of labelled equilateral triangles.  
C are arcs of the equator that is made up of collinear configurations. 
This splits the triangleland shape sphere into two hemispheres of opposite orientation (clockwise and anticlockwise labelled 
triangles, as in Fig 5c). 
Then it is clear that mirror-image-identification of shapes in space becomes, in the triangleland shape space, reflection in about the equator, 
a concept which immediately generalizes to $N$ particles if reinterpreted as the complex conjugation operation.
This produces the hemisphere with edge, $\mathbb{H}^2$ [Fig \ref{Fig4}c)], the cone over this then is $\mathbb{R}^3_0$: the half-space with edge.  
The I are bimeridians of isoscelesness with respect to the 3 possible clusterings (i.e. choices of apex particle and base pair).   
Each of these separates the triangleland shape sphere into hemispheres of right and left slanting triangles with respect 
to that choice of clustering [Fig 5b)].
The R are bimeridians of regularness (equality of the 2 partial moments of inertia of the each of the possible 
2 constituent subystems: base pair and apex particle.)
Each of these separated the triangleland shape sphere into hemispheres of sharp and flat triangles with respect to that 
choice of clustering [Fig 5a)].  
The M are merger points: where one particle lies at the centre of mass of the other two. 
S denotes spurious points, which lie at the intersection of R and C but have no further notable properties (unlike the D, M or E points 
that lie on the other intersections).

Indistinguishability then involves quotienting out $S_3 \,\,\ \widetilde{=}\,\, \mathbb{D}_3$ (permutations of the 
particles, isomorphic to the dihaedral group acting on the triangle of double collision points D and 
distinguishing the 2 hemispheres) [Fig 4e) and 4i)].  
If mirror-image identification is absent, rather one quotients out $A_3 \,\,\widetilde{=}\,\, \mathbb{Z}_3$ (even permutations 
of the particles, isomorphic to the cyclic group acting on the triangle of double collision points D)  [Fig 4d) and 4h)].

%FFFFFFFFFFFFFFFFFFFFFFFFFFFFFFFFFFFFFFFFFFFFFFFFFFFFFFFFFFFFFFFFFFFFFFFFFFFFFFFFFFFFFFFFFFFFFFFFFFFFFFF
{            \begin{figure}[ht]
\centering
\includegraphics[width=1.0\textwidth]{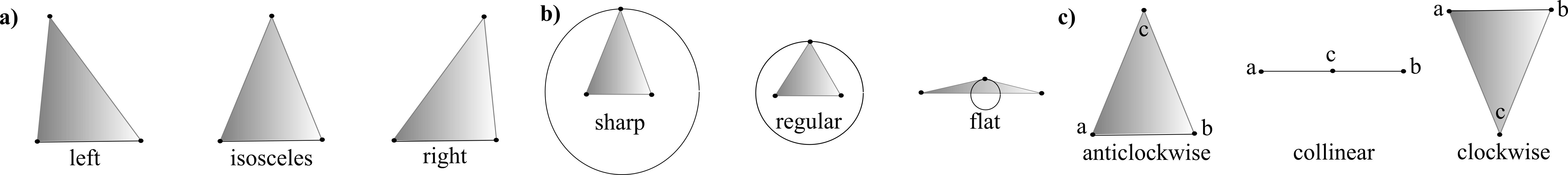}
\caption[Text der im Bilderverzeichnis auftaucht]{        \footnotesize{The most physically meaningful 
great circles on the triangleland shape correspond to the isosceles, regular and collinear triangles.
These respectively divide the shape sphere into hemispheres of right and left triangles, sharp and flat 
triangles, and anticlockwise and clockwise triangles.} }
\label{Fig6} \end{figure}          }
%FFFFFFFFFFFFFFFFFFFFFFFFFFFFFFFFFFFFFFFFFFFFFFFFFFFFFFFFFFFFFFFFFFFFFFFFFFFFFFFFFFFFFFFFFFFFFFFFFFFFFFFF

For later comparison with quadrilateralland,

\noindent 1) useful redundant coordinates covering a surrounding Euclidean space 
(here also the relational space) are now the complicated combinations of the two Jacobi vectors, namely  
Dragt coordinates 
\beq
\mbox{dra}_1 = 2\,{\bn}_1 \cdot {\bn}_2 \mbox{ } \mbox{ } , \mbox{ } 
\mbox{dra}_2 = 2\{{\bn}_1 \cr {\bn}_2\}_3   \mbox{ } \mbox{ } , \mbox{ }
\mbox{dra}_3 = |\mn_2|^2 - |\mn_1^2| \mbox{ } .
\eeq
These are, respectively \cite{+tri}, the anisoscelesness of the labelled triangle, four times the mass-weighted 
area of the triangle per unit moment of inertia and the ellipticity of the two partial moments of inertia.
The area Dragt coordinate is a democratic invariant and is useable as a measure of uniformity \cite{QShape}, 
its modulus running from maximal value at the most uniform configuration (the equilateral triangle) to minimal 
value for the collinear configurations. 
The on-sphere condition is then $\sum_{\Delta = 1}^3\mbox{dra}_{\Delta}\mbox{}^2 = 1$.\footnote{$\Delta$ is 
%%%%%%%%%%%%%%%%%%%%%%%%%%%%%%%%%%%%%%%%%%%%%%%%%%%%%%%%%%%%%%%%%%%%%%%%%%%%%%%%%%%%%%%%%%%%%%%%%%%%%%%%%%%%%%%%
an index running from 1 to 3 for triangleland and from 1 to 6 for quadrilateralland, whilst $\delta$ is an 
index running from 1 to 2 for triangleland and from 1 to 5 for quadrilateralland.}
%%%%%%%%%%%%%%%%%%%%%%%%%%%%%%%%%%%%%%%%%%%%%%%%%%%%%%%%%%%%%%%%%%%%%%%%%%%%%%%%%%%%%%%%%%%%%%%%%%%%%%%%%%%%%%
%
These combinations appearing as surrounding Cartesian coordinates is much less obvious than the $\mn^e$ appearing in the 
same role for 4-stop metroland.  
These combinations arise from the sequence $\mathbb{R}^4 \longrightarrow \mathbb{S}^3 \longrightarrow 
\mathbb{S}^2 \longrightarrow \mathbb{R}^3$, where the first map is an obvious on-sphere condition 
for preshape space within relative space, the second is mathematically the Hopf map and here sends preshape space to 
shape space, and the third is an obvious coning that here sends shape space to relative space.  

\noindent 2) 
One sometimes also swaps the dra$_{2}$ = demo(3) (the democratic invariant shape quantity for triangleland) for the scale variable $I$ in the non-normalized version of the coordinates to obtain the \{$I$, aniso, ellip\} system.  

\noindent 3) 
Then a simple linear recombination of this is \{$I_1$, $I_2$, aniso\}, i.e. the two partial moments of inertia and 
the dot product of the two Jacobi vectors.  
This is in turn closely related \cite{08I} to the parabolic coordinates on the flat $\mathbb{R}^3$ conformal to the 
triangleland relational space, which are $\{I_1, I_2, \Phi\}$.  

\noindent 4) Useful intrinsic coordinates are $\Theta$ and $\Phi$.   
These are again spherical polars, but their meaning in terms of the relative particle 
cluster separations is rather different.  
The interpretation of the azimuthal angle is now 
\beq
\Theta = 2\,\mbox{arctan}(\rho_2/\rho_1)
\label{TriAzi} \mbox{ } ,  
\eeq 
and that of the polar angle $\Phi$ is as in Fig 1d).

%======================================================================================================
\section{Quadrilateralland}
%======================================================================================================

%======================================================================================================
\subsection{Overview of configuration spaces for quadrilateralland}
%======================================================================================================

%FFFFFFFFFFFFFFFFFFFFFFFFFFFFFFFFFFFFFFFFFFFFFFFFFFFFFFFFFFFFFFFFFFFFFFFFFFFFFFFFFFFFFFFFFFFFFFFFFFFFFFF
{            \begin{figure}[ht]
\centering
\includegraphics[width=0.8\textwidth]{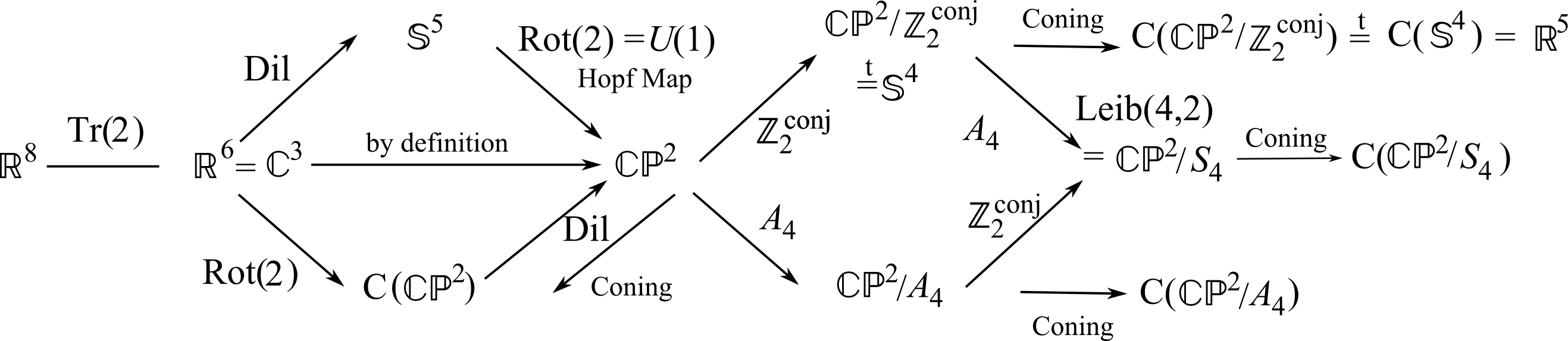}
\caption[Text der im Bilderverzeichnis auftaucht]{        \footnotesize{The sequence of configuration spaces 
for quadrilateralland.  
$\stackrel{t}{=}$ denotes equality at the topological level.  } }
\label{Fig5} \end{figure}          }
%FFFFFFFFFFFFFFFFFFFFFFFFFFFFFFFFFFFFFFFFFFFFFFFFFFFFFFFFFFFFFFFFFFFFFFFFFFFFFFFFFFFFFFFFFFFFFFFFFFFFFFFF

\noindent N.B. this is harder to visualise than the previous subsections' shape spaces, due to greater dimensionality 
as well as greater geometrical complexity and a larger hierarchy of special regions of the various possible codimensions. 
Geometrical detail of this space is in part built up in later Subsections.

Firstly, at the topological level, in the distinguishable particle plain shape case, the shape space is a $\mathbb{CP}^2$ 
decorated by a net of 6 $\mathbb{S}^2$ trianglelands (in each case with one vertex being a double collision 
corresponding to the $ \mbox{\Large (} \stackrel{\mbox{\scriptsize 4}}
                                                {\mbox{\scriptsize 2}}\mbox{\Large )}$  pairs of particles coinciding); each of these is an $\mathbb{S}^2$. 
There is also a net of 3-stop metrolands with one being a D point on an $\mathbb{RP}^n$ mirror image
identified by rotation in 2-$d$, with further special T and DD points the
usual 4 and 3 for a mirror-image identified such, and in the same pattern as in Fig 3.  
In the distinguishable particle mirror image identified case, the shape space is a 
$\mathbb{CP}^2/\mathbb{Z}_2^{\sc\so\sn\sj}$ (see \cite{WCP} for similar quotients 
but with a different meaning to the $\mathbb{Z}_2$ action) decorated by a net of 6
$\mathbb{RP}^2$큦 trianglelands.
In the indistinguishable particle plain shape case, the shape space is $\mathbb{CP}^2/A_4$, with the triangular configurations 
feeling only $A_3$. 
In the indistinguishable particle mirror-image-identified case, the shape space is $\mathbb{CP}^2/S_4$, with
the triangular configurations feeling only $S_3$ but the $\mathbb{RP}^2$ of collinear configurations feeling the full $S_4$.

Secondly, at the metric level, collinearities become meaningful. 
These form $\mathbb{RP}^2$ in the distinguishable particle cases and $\mathbb{RP}^2/A_4$ in the 
indistinguishable particle cases.  
Here is a demonstration that $\mathbb{RP}^{N - 2}$ plays this role within the general $N$-a-gonland $\mathbb{CP}^{N - 2}$ case.
Collinear configurations involve the relative angle coordinates being 0 or 
multiples of $\pi$, by which the complex projective space definition collapses to the 
real projective space definition (for which the ${\cal R}^e$ are Beltrami coordinates).
Moreover, we know that abcd... can be rotated via the second dimension into ...dcba (and that there 
is no further such identification) and so the real submanifold of collinear configurations is $\mathbb{RP}^{N - 2}$.  
The above spaces of collinearity are in each case like an equator in each case, e.g. separating two  $\mathbb{S}^4$'s in
$\mathbb{E}^5$ in the first case (see below for why the two halves are, topologically, $\mathbb{S}^4$ and for further 
issues of the geometry involved).

Finally, there are also 6, 3, 2 and 1 distinguishable labelled squares in each case.
This motivates a new action on quadrilateralland, in which e.g. the left-most particle is preserved and
the other 3 are permuted or just evenly-permuted.

It is also easy to write conditions in these coordinates for rectangles, kites, trapezia, rhombi... but
these are less meaningful 1) from a mathematical perspective (e.g. they are not topologically defined).   
2) From a physical perspective (the square is additionally a configuration for which various 
notions of uniformity are maximal). 
However, squares are not the only notion of maximal uniformity by \cite{QShape}'s the $s_6$ =demo(4) 
quantifier described in Sec 5.2; I further investigate this quantifier of uniformity using the present paper's 
coordinate systems in   Sec 5.5. 
Parallelograms also play a role in Figure 7.

%==============================================================================================================
\subsection{Useful coordinate systems for quadrilateralland.} 
%==============================================================================================================

Further useful coordinate systems for quadrilateralland are as follows [these parallel the same 1) to 4) labels as in Sec 4].

\noindent
1) \{$s_{\Delta}$\} are a redundant set of six shape coordinates (see \cite{QShape} for their explicit forms), 
which are the quadrilateralland analogue of triangleland's Dragt coordinates according to the construction in 
\cite{LR95, QShape}.    

\mbox{ }

\noindent 
2) However, for quadrilateralland, swapping one of the $s_{\Delta}$ [$s_6$ = demo($4$), a democracy invariant and 
proportional to the square root of the sum of the squares of the mass-weighted areas per unit MOI of 
the three coarse-graining triangles and parallelogram of Fig \ref{FigX}b-d) and g-i)] for a scale variable to 
form the coordinate system \{$I$, $s_{\delta}$\} turns out to give a more useful set \cite{QShape}.

\mbox{ } 

\noindent 
3) Kuiper's coordinates are then a simple linear combination of 2) (mixing the $I$, $s_1$ and $s_5$ coordinates).   
These consist of all the possible inner products between pairs of Jacobi vectors, i.e. 3 magnitudes of Jacobi vectors per unit MOI 
$N^e$, alongside 3  $\bn_f\cdot\bn_g$ ($e \neq f$), which are very closely related to the three relative angles.  
As such, they are, firstly, very much an extension of the parabolic coordinates for the conformally-related flat 
$\mathbb{R}^3$ of triangleland \cite{08I}.  
Secondly, in the quadrilateralland setting, they are a clean split into 3 pure relative angles (of which any 2 are independent 
and interpretable as the anioscelesnesses of the coarse-graining triangles or arhombicness of the coarse-graining parallelogram) and 3 magnitudes 
(supporting 2 independent non-angular ratios).
I therefore denote this coordinate system by \{$N^e$, aniso($e$)\}.  
Thirdly, whilst they clearly contain 2 redundancies, they are fully democratic in relation to the constituent 
Jacobi vectors and coarse-graining triangles/parallelogram made from pairs of them.

%FFFFFFFFFFFFFFFFFFFFFFFFFFFFFFFFFFFFFFFFFFFFFFFFFFFFFFFFFFFFFFFFFFFFFFFFFFFFFFFFFFFFFFFFFFFFFFFFFFFFFFFFFFFFFFFFFF
{            \begin{figure}[ht]
\centering
\includegraphics[width=0.8  \textwidth]{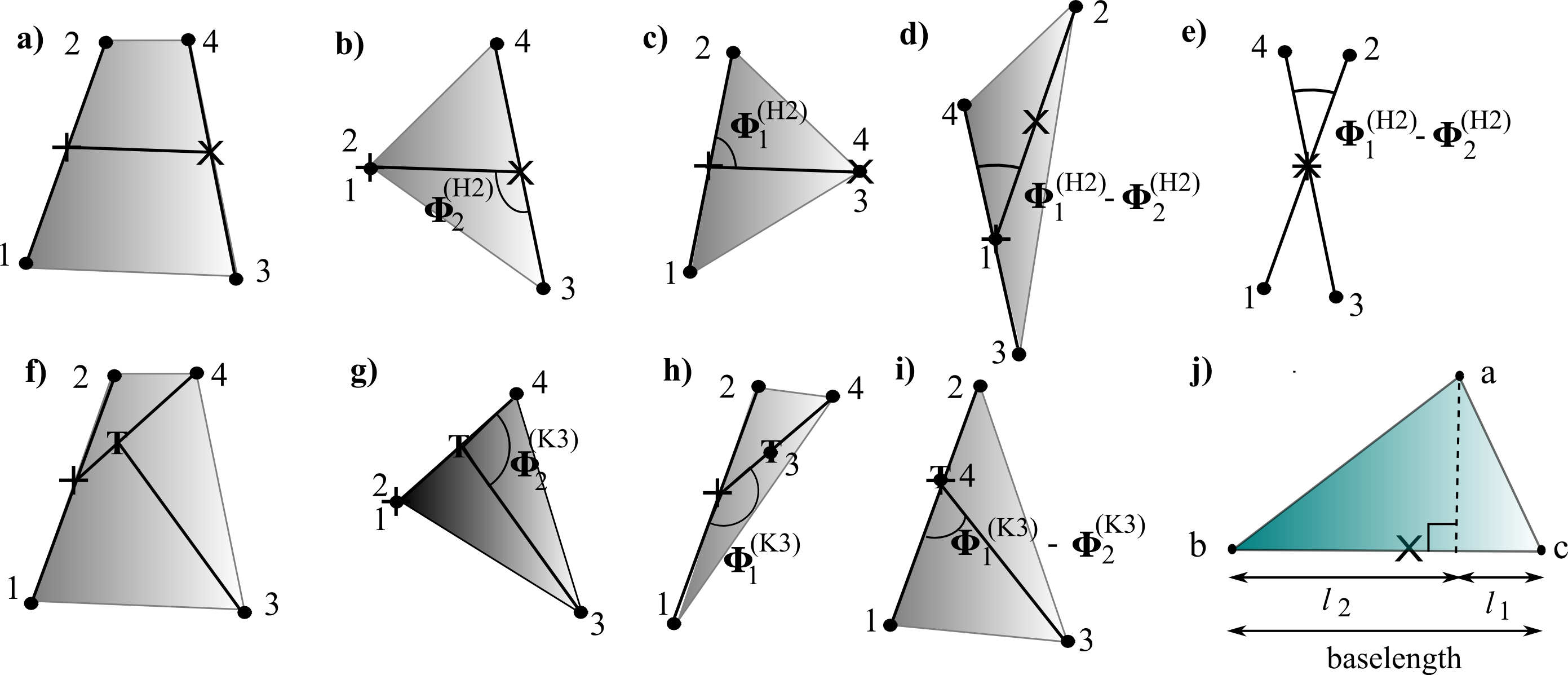}
\caption[Text der im Bilderverzeichnis auftaucht]{        \footnotesize{Figure of the coarse-graining triangles.  
a) For Jacobi H-coordinates coordinates for a given quadrilateral,    
collapsing each of $\rho_1^{(\tH 2)}$, $\rho_2^{(\tH 2)}$ and $\rho_3^{(\tH 2)}$ in turn gives the coarse-graining 
triangles b), c) and d) \cite{QShape}. 
[One can use d) to interpret the associated anisoscelesness, but the true physical 
situation is that of e) i.e. a parallelogram.]  
Moreover, whilst both the triangles and the parallelograms have spheres for their reduced configuration spaces, 
the detail of how the sphere of parallelograms is decorated differs from how the sphere of triangles is decorated, 
so that one really needs to use e) itself. 
Then the analogue of anisoscelessness here 
is arhombicness, i.e. departure from the rightness of the angle of intersection of the diagonals that characterizes a rhombus.]
If one uses Jacobi K-coordinates instead [f), new to this paper], then 
collapsing each of $\rho_1^{(\tK 3)}$, $\rho_2^{(\tK 3)}$ and $\rho_3^{(\tK 3)}$ in turn gives 
the coarse-graining triangles g), h) and i).
j) The anisoscelesness per unit base length is the amount by which the perpendicular to the base fails to bisect 
it is $l_1 - l_2$ for $l_1$ and $l_2$ as indicated.}         }
\label{FigX}\end{figure}          }
%FFFFFFFFFFFFFFFFFFFFFFFFFFFFFFFFFFFFFFFFFFFFFFFFFFFFFFFFFFFFFFFFFFFFFFFFFFFFFFFFFFFFFFFFFFFFFFFFFFFFFFFFFFFFFFFFFF

%\mbox{ }

\noindent 4) Useful intrinsic coordinates (which extend the spherical coordinates on the triangleland and 4-stop metroland 
shape spheres) are the Gibbons--Pope type coordinates \{$\chi$, $\beta$, $\phi$, $\psi$\} are also useful in this paper.  
The coordinate ranges are $0 \leq \chi \leq \pi/2$, 
                 $0 \leq \beta \leq \pi$, 
                 $0 \leq \phi \leq 2\pi$ (a reasonable range redefinition since it is the third relative angle), and
                 $0 \leq \psi \leq 4\pi$.
These are related to the bipolar form of the Fubini--Study coordinates by
\beq
\psi^{\prime} = - \{\Phi_1 + \Phi_2\} \mbox{ } , \mbox{ } \mbox{ } \phi^{\prime} = \Phi_2 -\Phi_1, 
\eeq
with then $\psi = - \psi^{\prime}$ (measured in the opposite direction to match Gibbons--Pope's convention) 
and $\phi$ is taken to cover the coordinate range $0$ to $2\pi$, which is comeasurate with it itself being the 
third relative angle between the Jacobi vectors involved.
However now in each of the conventions I use for H and K coordinates, a different interpretation is to be attached to these last two 
formulae in terms of the $\rho^e$.
For H-coordinates in my convention, 
\beq
\beta = 2\,\mbox{arctan}\,(\rho_2/\rho_1)  \mbox{ } , \mbox{ } \mbox{ } \chi = \mbox{arctan}\big(\sqrt{\rho_1\mbox{}^2 + \rho_2\mbox{}^2}/\rho_3\big) \mbox{ } ,
\eeq
whilst for K-coordinates in my convention, 
\beq
\beta = 2\,\mbox{arctan}\,(\rho_1/\rho_2)  \mbox{ } , \mbox{ } \mbox{ } \chi = \mbox{arctan}\big(\sqrt{\rho_1\mbox{}^2 + 
\rho_2\mbox{}^2}/\rho_3\big) \mbox{ } .
\eeq
\mbox{ } \mbox{ }  
By their ranges, $\beta$ and $\phi$ parallel azimuthal and polar coordinates on the sphere.
[In fact, $\beta$, $\phi$ and $\psi$ take the form of Euler angles on $SU(2)$, with the remaining coordinate $\chi$ playing the 
role of a compactified radius.]   
Now, in the quadrilateralland interpretation, $\beta$ has the same mathematical form as triangleland's 
azimuthal coordinate $\Theta$ (\ref{TriAzi}).
Additionally, $\chi$ parallels 4-stop metroland's azimuthal coordinate $\theta$ [the first equation in 
(\ref{4StopPolars})], except that it is over half of the range of that, reflecting that the collinear 1234 and 
4321 orientations have to be the same due to the existence of the second dimension via which one is rotateable 
into the other.

The Gibbons--Pope type coordinates have the following quadrilateralland interpretations.   
In an H-clustering, the $\phi$ is the difference of the relative angles [see Fig 2e)], 
so the associated momentum represents a counter-rotation of the two constituent subsystems ($\times$ relative to \{12\} and 
+ relative to \{34\}).       
The $\psi$ is minus the sum of the relative angles, so the associated momentum represents a co-rotation of these two 
constituent subsystems (with counter-rotation in $\times$ relative to $+$ so as to preserve the overall zero angular 
momentum condition).  
The $\beta$ is a measure of contents inhomogeneity of the model universe: the ratio of the sizes of the 2 constituent subclusters.  
Finally, the $\chi$ is a measure of the selected subsystems' sizes relative 
to that of the whole-universe model.  
These last two are conjugate to quantities that involve relative dilational momenta in addition to relative angular momenta.

On the other hand, in a K-clustering, the $\phi$ is the difference of the two relative angles [see Fig 2f)], but by the 
directions these are measured in, the associated momentum now represents a co-rotation of the two constituent subsystems (which are now \{12\} relative to 4 and \{+4\} relative to 3, and with  counter-rotation in T relative to $+$ so as to preserve the overall zero 
angular momentum condition).   
The $\psi$ is minus the sum of the two relative angles, so the associated momentum represents a counter-rotation of these two 
constituent subsystems. 
The $\beta$ is now a comparer between the sizes of the \{12\} subcluster and the separation between the non-triple cluster 
particle 3 and T.
Finally, the $\chi$ is a comparer between the sizes of the above two contents of the universe (\{12\} and \{T3\})  on the 
one hand, and the separation between them on the other hand (\{4+\}, which is a measure of separation of \{12\} and \{T3\}). 
These last two are again conjugate to quantities that involve relative dilational momenta in addition to relative angular momenta.
Note how in both H and K cases, the Gibbons--Pope type coordinates under the quadrilateralland interpretation involve a split 
into two pure relative angles and two pure non-angular ratios of magnitudes.

In Gibbons--Pope type coordinates, the Fubini--Study metric then takes the form  
\beq
\d s^2 = \d\chi^2 + 
\mbox{sin}^2\chi\big\{\d \beta^2 + \mbox{cos}^2\chi\{\d\phi^2 + \d\psi^2 + 
2\,\mbox{cos}\,\beta\, \d\phi\d\psi\} + \mbox{sin}^2\chi\,\mbox{sin}^2\beta \,\d\phi^2 
\big\}/4 \mbox{ } , 
\eeq
leading to a kinetic term
\beq
\fT = \dot{\chi}^2/2 + 
\mbox{sin}^2\chi\big\{\dot{\beta}^2 + \mbox{cos}^2\chi\{\dot{\phi}^2 + \dot{\psi}^2 + 
2\,\mbox{cos}\,\beta\, \dot{\phi}\dot{\psi}\} + \mbox{sin}^2\chi\,\mbox{sin}^2\beta \, \dot{\phi}^2 
\big\}/8 \mbox{ } .   
\eeq

%================================================================================================
\subsection{The inclusion of trianglelands and 4-stop metroland within quadrilateralland}
%================================================================================================

In Gibbons--Pope type coordinates based on the Jacobi H,  
when $\rho_3 = 0$, $\chi = \pi/2$ and the metric reduces to 
\beq
\d s^2 = \{1/2\}^2\{\d \beta^2 + \mbox{sin}^2\beta\,\d\phi^2\}
\eeq
i.e. a sphere of radius 1/2, which corresponds to the conformally-untransformed $\mathbb{CP}^1$.  
When $\rho_2$ = 0, $\beta = 0$ and the metric reduces to 
\beq
\d s^2 = \{1/2\}^2\{\d \Theta_1^2 + \mbox{sin}^2\Theta_1\d\Phi_1^2\}
\eeq
for $\Theta_1 = 2\chi$ having the correct coordinate range for an azimuthal angle.  
Finally, when $\rho_1$ = 0, $\beta = 0$ and the metric reduces to 
\beq
\d s^2 = \{1/2\}^2\{\d \Theta_2^2 + \mbox{sin}^2\Theta_2\d\Phi_2^2\}
\eeq
for $\Theta_2 = 2\chi$ again having the correct coordinate range for an azimuthal angle.   
The first two of these spheres are a triangleland shape sphere included within quadrilateralland as per 
Sec 5.1.  
The first is for \{12\}, 4 and 3 as the particles.  
The second is for 1, 2 and 34 as the particles. 
The third of these spheres corresponds, rather, to a merger, of + and $\times$, i.e. a merger of type 
$\mM^{\sD\sD}$ -- the space of parallelograms labelled as in Fig 7e).

In Gibbons--Pope type coordinates based on the Jacobi K, when $\rho_3 = 0$, $\chi = \pi/2$ and the metric reduces to 
\beq
\d s^2 = \{1/2\}^2\{\d \beta^2 + \mbox{sin}^2\beta\,\d\phi^2\}
\eeq
i.e. a sphere of radius 1/2, which corresponds to the conformally-untransformed $\mathbb{CP}^1$.  
When $\rho_2$ = 0, $\beta = 0$ and the metric reduces to 
\beq
\d s^2 = \{1/2\}^2\{\d \Theta_1^2 + \mbox{sin}^2\Theta_1\d\Phi_1^2\}
\eeq
for $\Theta_1 = 2\chi$ again having the correct coordinate range for an azimuthal angle.  
Finally, when $\rho_1$ = 0, $\beta = 0$ and the metric reduces to 
\beq
\d s^2 = \{1/2\}^2\{\d \Theta_2^2 + \mbox{sin}^2\Theta_2\d\Phi_2^2\}
\eeq
for $\Theta_2 = 2\chi$ yet again having the correct coordinate range for an azimuthal angle.   
The second of these is a triangleland shape space sphere with 12, 4 and 3 as the particles. 
The first and third correspond rather to mergers, of +, T and 4 in the first case 
(of type $\mM^{\sT} \cap \mM^{\sD\sD}$), and of T and 3 in the second case (of type $\mM^{\sT}$).

In Kuiper coordinates, each of the three on-$\mathbb{S}^2$ conditions for whichever of H or K coordinates involves 
losing one magnitude and two inner products.  
Thus the survivors are two magnitudes and one inner product (closely related to parabolic coordinates and  
linearly combineable to form the \{$I$, aniso, ellip\} system, see Sec 4).  
If we recombine the Kuiper coordinates to form the \{$I, s_{\delta}$\} system, and swap the $I$ for the $s_6$ = demo(4), 
then the survivors of the procedure are the Dragt coordinates (since the procedure kills two of the three area contributions 
to the $s_6$).  
This gives another sense in which the \{$s_\Delta$\} system is a natural extension of the Dragt system.

%================================================================================================
\subsection{The split into hemi-$\mathbb{CP}^2$'s of oriented quadrilaterals}
%================================================================================================

{\bf Definition}: The Veronese surface $V$ \cite{Veronese} is the space of conics through a point (parallel to how 
a projective space is a set of lines through a point).  

\mbox{ }

\noindent {\bf Kuiper's Theorem} i) The map  

\beq
\eta: \stackrel{\mbox{$\mathbb{CP}^2$}}{ (\mz_1, \mz_2, \mz_3) }                  
\mbox{ } \mbox{ } \mbox{ }
      \stackrel{\mbox{$\longrightarrow$}}{   \stackrel{}{\mbox{$\longmapsto$}}   }  
\mbox{ } \mbox{ } \mbox{ }
      \stackrel{\mbox{$\mathbb{E}^5$}}{(|\mz_1|^2, |\mz_2|^2,|\mz_3|^2, 
\{\mz_2\bar{z}_3 + \mz_3\bar{z}_2\}/2, \{\mz_3\bar{z}_1 + \mz_1\bar{z}_3\}/2, \{\mz_1\bar{z}_2 + \mz_2\bar{z}_1\}/2)}
\eeq  
induces a piecewise smooth embedding of $\mathbb{CP}^2/\mathbb{Z}_2^{\sc\so\sn\sj}$  onto the 
boundary of the convex hull of the Veronese surface $V$ in $\mathbb{E}^5$, which moreover has the right 
properties to be the usual smooth 4-sphere \cite{Kuiper}. 

%\mbox{ } 

\noindent N.B. this is at the topological level; it clearly cannot extend to the metric level by 
a mismatch in numbers of Killing vectors ($\mathbb{S}^4$ with the standard spherical metric has 10 
whilst $\mathbb{CP}^2$ equipped with the Fubini--Study metric has 8).    

\mbox{ } 

\noindent Restricting $\{{\mz}_1, {\mz}_2, {\mz}_3\}$ to the real line corresponds in the quadrilateralland interpretation 
to considering the collinear configurations, which constitute a $\mathbb{RP}^2$ space as per Sec 5.1  
Moreover the above embedding sends this onto the Veronese surface $V$ itself.  
Proving this proceeds via establishing that, as well as the on-$\mathbb{S}^5$ condition $N_1 + N_2 + N_3 = 1$,   
a second restriction holds, which in our quadrilateralland interpretation, reads 
$\sum_{e = 1}^n \mbox{aniso}(e)^2N^e - 4N_1N_2N_3 + \mbox{aniso}(1)\,\mbox{aniso}(2)\, 
\mbox{aniso}(3) = 0$. 
(Knowledge of this restriction should also be useful in kinematical quantization \cite{Isham84}, and it is clearer in the \{$N^e$, aniso($e$)\} 
system, which is both the quadrilateralland interpretation of Kuiper's redundant coordinates and a simple linear recombination of 
the  \{$I$,$s_{\delta}$\} coordinates obtained in \cite{QShape}, than in these other coordinates themselves.)  

\mbox{ }

\noindent 
{\bf Another form for Kuiper's theorem} \cite{Kuiper}.
Moreover, $\mathbb{CP}^2$ itself is topologically a double covering of $\mathbb{S}^4$ branched along the $\mathbb{RP}^2$ 
of collinearities which itself embeds onto $\mathbb{E}^5$ to give the Veronese surface $V$. 

\mbox{ } 

\noindent Here, branching is meant in the sense familiar from the theory of Riemann surfaces \cite{Riemann}. 
Moreover,  the $\mathbb{RP}^2$ itself embeds {\sl non-smoothly} into the Veronese surface $V$.

\mbox{ }

\noindent
Then the quadrilateralland interpretation of these results is in direct analogy with the plain shapes case 
of triangleland consisting of two hemispheres of opposite orientation bounded by an equator circle of collinearity, 
the mirror-image-identified case then consisting of one half plus this collinear edge.  
Thus plain quadrilateralland큦 distinction between clockwise- and anticlockwise-oriented figures is strongly anchored 
to this geometrical split, with the collinear configurations lying at the boundary of this split.

%================================================================================================
\subsection{Notions of uniformity cast in Kuiper and Gibbons--Pope coordinates}
%================================================================================================

In this quadrilateralland case, there are then 3 squares in each hemi-$\mathbb{CP}^2$ as 
opposed to the single equilateral triangle in each hemisphere of triangleland; 
these are particularly uniform configurations.  
This reflects the presence of a further 3-fold symmetry in quadrilateralland; 
choosing to use indistinguishable particles then quotients this out.

As explained in detail in \cite{+tri}, the nontrivial democratic invariant demo(3) for 
triangleland is the mass-weighted area of the triangle per unit moment of inertia, and that for 
quadrilateralland, demo(4), is proportional to the square root of the sum of the mass-weighted 
areas of the coarse-graining triangles/parallelogram in Fig 7.  
Now, extremizing demo(3) invariant picks out the two labelled equilateral triangles, 
which is uniform in a very strong sense; on the other hand, extremizing demo(4) does pick out the six 
labelled squares, but nonuniquely -- one gets one extremal curve per hemi-$\mathbb{CP}^2$. 
Now, the present paper's choices of coordinate systems furthermore clarify the interpretation to be given to 
these two extremal curves.  
In $\{N^e$, aniso($e$)\} coordinates, this is given by aniso(1) = 0 = aniso(2) and 
$\left| \mbox{aniso(3)}\right|$ = 1,  i.e. (1)-isosceles, (2)-isosceles and maximally (3)-right or (3)-left i.e. 
(3)-collinear, with $N_3$ = 1/2 and $N_1$, $N_2$ varying (but such that the on-$\mathbb{S}^5$ condition $N_1 + N_2 + N_3 = 1$ holds).  
On the other hand, in  Gibbons--Pope type coordinates, the uniformity condition is $\phi = 0$, $\psi = 4\pi$, 
$\chi = \pi/4$ and $\beta$ free, i.e., in the H-coordinates case, freedom in the contents inhomogeneity i.e. size 
of subsystem 1 relative to the size of subsystem 2.
On the other hand, in the K-coordinates case, $\beta$ free signifies freedom in how tall one makes 
the selected \{12, 3\} cluster's triangle.

%===================================================================================================================
%===================================================================================================================
\section{Conclusion}
%===================================================================================================================
%===================================================================================================================

%===================================================================================================================
\subsection{Outline of the results so far}
%===================================================================================================================

Quadrilateralland's shape space is $\mathbb{CP}^2$ or some quotient of this by a discrete group. 
I considered this for plain and mirror-image identified choices of shapes, and for distinguishable and 
indistinguishable particles. 
In the case in which distinct particle masses do not naturally label the particles as distinct, the most 
Leibnizian shape space is Leib$(4,2) =\mathbb{CP}^2/S_4$.
I have also explained how trianglelands and 4-stop metrolands occur within quadrilateralland; 
these correspond at the level of space to double collision configurations and collinear configurations. 
At the level of configuration spaces, they correspond respectively to $\mathbb{S}^2$'s or quotients and 
to $\mathbb{RP}^2$'s or quotients.
I have also drawn attention to Kuiper's theorem as providing, in the present quadrilateralland context,  
the analogue of the 2 hemispheres of orientation separated by an equator of collinearity for triangleland 
and the codimension-1 embedding thereof into a surrounding Euclidean space.
This paper was also the first to present details of the Jacobi K-coordinate study of 4-stop metroland and 
quadrilateralland, and of the structure of the set of mergers for 4-stop metroland and quadrilateralland.

I identified the following coordinate systems as useful for the study of quadrilateralland. 
 
\mbox{ } 
 
\noindent 1) Kuiper coordinates (a redundant presentation on $\mathbb{C}^3$).  
These consist of magnitudes of the 3 mass-weighted Jacobi vectors and the 3 inner products between 
these vectors.  
These are then physically interpreted as the three partial moments of inertia of the system and the three 
anisoscelesneses of Fig \ref{FigX}'s coarse-graining triangles (with one substituted for an arhombicness 
of a parallelogram in the H case) obtained  from the quadrilateral 
by striking out each Jacobi vector in turn.   
These are closely related to triangleland's conformally-related flat space's parabolic coordinates: two partial 
moments of inertia and the relative angle between the two Jacobi vectors.  

\mbox{ } 

\noindent 2) Gibbons--Pope type coordinates, which, for quadrilateralland in an H-clustering, are to be identified as 
a difference of relative angles,  minus the sum of the relative angles, a measure of contents inhomogeneity and a 
measure of the selected subsystems' sizes relative to that of the whole-universe model.  
In K-coordinates, the last two are, rather, i) a comparer between the sizes of the \{12\} subcluster and the separation between the non-triple cluster particle 3 and T, and ii) a comparer between the sizes of the above two contents of the universe (\{12\} and \{T3\})  on the 
one hand, and the separation between them on the other hand (\{4+\}, which is a measure of separation of \{12\} and \{T3\}). 

In H-coordinates, the momentum associated with the difference of relative angles represents a counter-rotation of the 
two constituent subsystems ($\times$ relative to \{12\} and + relative to \{34\}), while the momentum associated with 
the sum of relative angles represents a co-rotation of these two constituent subsystems (with counter-rotation in 
$\times$ relative to $+$ so as to preserve the overall zero angular momentum condition).  
On the other hand, in K-coordinates, the momentum associated with a difference of relative angles represents a 
co-rotation of the two constituent subsystems (which are now \{12\} relative to 4 and \{+4\} relative to 3, and with 
counter-rotation in T relative to $+$ so as to preserve the overall zero angular momentum condition), whilst that 
associated with the sum of relative angles  represents a counter-rotation of these two constituent subsystems. 
These momenta will turn out to play an important role in the study of conserved quantities in quadrilateralland 
\cite{QCons}.  

\mbox{ }  
 
\noindent Square configurations and a weaker criterion of uniform states based on the extremization of the 
sum of the squares of the mass-weighted areas per unit MOI of the three coarse-graining triangles (or two triangles 
and one parallelogram) were furtherly  
understood using this paper's coordinate systems.

%==========================================================================================
\subsection{Generalizations of this paper}
%==========================================================================================

%FFFFFFFFFFFFFFFFFFFFFFFFFFFFFFFFFFFFFFFFFFFFFFFFFFFFFFFFFFFFFFFFFFFFFFFFFFFFFFFFFFFFFFFFFFFFFFFFFFFFFFF
{            \begin{figure}[ht]
\centering
\includegraphics[width=0.75\textwidth]{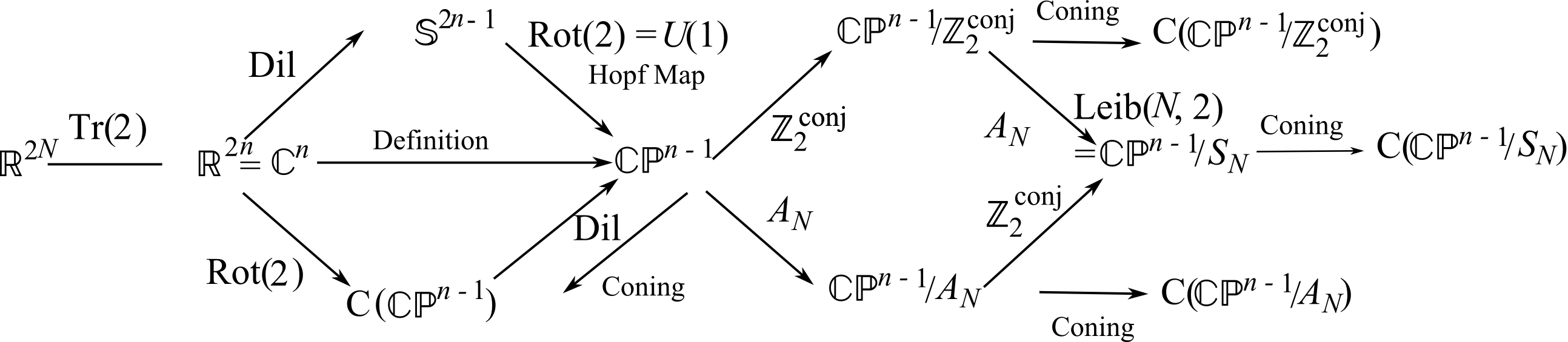}
\caption[Text der im Bilderverzeichnis auftaucht]{        \footnotesize{The sequence of configuration 
spaces for $N$-a-gonland.} }
\label{Fig7} \end{figure}          }
%FFFFFFFFFFFFFFFFFFFFFFFFFFFFFFFFFFFFFFFFFFFFFFFFFFFFFFFFFFFFFFFFFFFFFFFFFFFFFFFFFFFFFFFFFFFFFFFFFFFFFFFF

\noindent 
Extending the Dragt/parabolic/shape/Kuiper type of redundant coordinates is itself straightforward, 
though it is not clear the extent to which the resulting coordinates will retain usefulness for the 
study of each $\mathbb{CP}^{N - 2}$.  
Certainly the number of Kuiper-type coordinates (based on inner products of pairs of Jacobi vectors, 
of which there are $N\{N - 1\}/2$) further grows away from 2$N$ = dim($\mathbb{CP}^{\sN}$) as $N$ gets larger.
The present paper also finds a surrounding space of just {\sl one} dimension more for $\mathbb{CP}^2$.  
The $N$-a-gonland significance of two half-spaces of different orientation separated by an orientationless manifold of 
collinearities gives reason for double covers to the $N - 2 \geq 3$ $\mathbb{CP}^{N - 2}$ spaces to 
exist for all $N$.  
However, there is no known guarantee that these will involve geometrical entities as simple as or tractable as 
quadrilateralland's $\mathbb{S}^4$ for the half-spaces, or of the Veronese surface $V$ as the place of branching.  
However, one does have the simple argument of Sec 5.1 that the manifold of collinearities within $N$-a-gonland's 
$\mathbb{CP}^{N - 2}$ is $\mathbb{RP}^{N - 2}$, so at least that is a known and geometrically-simple result 
for the structure of the general $N$-a-gonland.
Whether the intrinsic Gibbons--Pope type coordinates can be extended to $N$-a-gonland in a way that maintains their 
usefulness in characterizing conserved quantities \cite{QCons} and via 
separating the free-potential time-independent Schr\"{o}dinger equation, remains to be seen.

As regards the Quantum Information Theory counterpart of this work, one possible usefulness of representing qtrit states as quadrilaterals 
is via the convenience of having a 2-$d$ graphical representation, which, moereover, remains 2-$d$ as one passes to the study of q$n$its.  
In this picture, the relation between the 3 included $SU(2)$ ladders and the 3 coarse-graining triangles (or two triangles and one parallelogram) is that there are 3 constituent (overlapping) qbits in a qtrit.
I leave what the Gibbons--Pope and Kuiper coordinates (and the associated democratic invariant) signify in the study of Qutrits as an 
interesting open question in parallel to the present paper.

%===================================================================================================================
\subsection{RPM's and Problem of Time strategies}
%===================================================================================================================

Some of the strategies toward resolving the Problem of Time in Quantum Gravity modellable by RPM's are as 
follows \cite{FileR}.  
%
%Pageref.  

\mbox{ }

\noindent A) Perhaps one has slow heavy `$h$'  variables that provide an approximate timestandard with 
respect to which the other fast light `$l$' degrees of freedom evolve \cite{Semi, HallHaw, K92, Kieferbook}.  
In the Halliwell--Hawking \cite{HallHaw} scheme for GR Quantum Cosmology, $h$ is scale (and homogeneous 
matter modes) and $l$ are small inhomogeneities.
Thus the scale--shape split of scaled RPM's afford a tighter parallel of this \cite{MGM, SemiclIII, Forth} than 
pure-shape RPM's.    
The semiclassical approach involves firstly making the Born--Oppenheimer ansatz $\Psi(h, l) = \psi(\mh)
|\chi(h, l) \rangle$ and the WKB ansatz $\psi(h) = \mbox{exp}(iW(h)/\hbar)$.  
Secondly, one forms the $h$-equation ($\langle\chi| \hat{H} \Psi = 0$ for RPM's), which, under a number 
of simplifications, yields a Hamilton--Jacobi\footnote{For simplicity, I 
%%%%%%%%%%%%%%%%%%%%%%%%%%%%%%%%%%%%%%%%%%%%%%%%%%%%%%%%%%%%%%%%%%%%%%%%%%%%%%%%%%%%%%%%%%%%%%%%%%%%%%%%% 
present this in the case of 1 $h$ degree of freedom and with no linear constraints.} 
%%%%%%%%%%%%%%%%%%%%%%%%%%%%%%%%%%%%%%%%%%%%%%%%%%%%%%%%%%%%%%%%%%%%%%%%%%%%%%%%%%%%%%%%%%%%%%%%%%%%%%%%%
equation
\beq
\{\pa W/\pa h\}^2 = 2\{\fE - \fV(h)\} \mbox{ } 
\label{HamJac} 
\eeq
for $\fV(h)$ the $h$-part of the potential. 
Thirdly, one way of solving this is for an approximate emergent semiclassical time 
$t^{\se\sm} = t^{\se\sm}(h)$. 
Next, the $l$-equation $\{1 - |\chi\rangle\langle\chi|\}\hat{\H}\Psi = 0$ can be recast (modulo further 
approximations) into an emergent-time-dependent Schr\"{o}dinger equation for the $l$ degrees of freedom
\beq
i\hbar\pa|\chi\rangle/\pa t^{\te\tm}  = \widehat{H}_{l}|\chi\rangle \mbox{ } .  
\label{TDSE2}
\eeq
(Here the left-hand side arises from the cross-term $\pa_{h}|\chi\rangle\pa_{h}\psi$ and 
$\widehat{H}_{l}$ is the remaining surviving piece of $\widehat{H}$).  
Note that the working leading to such a time-dependent wave equation ceases to work in the absense of making 
the WKB ansatz and approximation, which, additionally, in the quantum-cosmological context, is not known to be a 
particularly strongly supported ansatz and approximation to make.    

\mbox{ }

\noindent B) A number of approaches take timelessness at face value. 
One considers only questions about the universe `being', rather than `becoming', a certain way.  
This has at least some practical limitations, but can address some questions of interest. 
As a first example, the {\it na\"{\i}ve Schr\"{o}dinger interpretation} \cite{HP86UW89} concerns the `being' 
probabilities for universe properties such as: what is the probability that the universe is large? 
Flat? 
Isotropic? 
Homogeneous?   
One obtains these via consideration of the probability that the universe belongs to region R of the 
configuration space that corresponds to a quantification of a particular such property, 
\beq 
\mbox{Prob(R)} \propto \int_{\sR}|\Psi|^2\d\Omega \mbox{ } , 
\label{NSI}
\eeq 
for $\d\Omega$ the configuration space volume element.
%
%This approach is termed `na\"{\i}ve' due to it not using any further features of the constraint 
%equations.  
%
As a second example, the {\it conditional probabilities interpretation} \cite{PW83} goes further by addressing 
conditioned questions of `being' such as `what is the probability that the universe is flat given that 
it is isotropic'?  
As a final example, {\it records theory} \cite{PW83, GMH, EOT, H99, Records} involves localized subconfigurations 
of a single instant.  
More concretely, it concerns whether these contain useable information, are correlated to each other, 
and a semblance of dynamics or history arises from this.  
This requires notions of localization in space and in configuration space as well as notions of 
information.  
RPM's are superior to minisuperspace for such a study as, firstly, they have a notion of localization 
in space. 
Secondly, they have more options for well-characterized localization in configuration space (i.e. of 
`distance between two shapes' \cite{FileR}) through their kinetic terms possessing positive-definite 
metrics.  

%\mbox{ }

\noindent C) Perhaps instead it is the histories that are primary ({\it histories theory} \cite{GMH, 
Hartle}).    

\mbox{ }

\noindent Combining A) to C) (for which RPM's are well-suited) is a particularly interesting prospect \cite{H03}, along the following lines 
(see also \cite{GMH, H99, H09, FileR} for further development of this).   
There is a records theory within histories theory.  
Histories decohereing is one possible way of obtaining a semiclassical regime in the first place, i.e. 
finding an underlying reason for the crucial WKB assumption without which the semiclassical approach does 
not work.    
What the records are will answer the also-elusive question of which degrees of freedom decohere 
which others in Quantum Cosmology.

\mbox{ }  

\noindent
(Observables-based approaches \cite{Rovellibook} can also be studied in the RPM arena, as can be 
quantum cosmologically aligned hidden time approaches for {\sl scaled} RPM's.)

%===================================================================================================================
\subsection{Use of regions of configuration space in Problem of Time approaches}
%===================================================================================================================

%FFFFFFFFFFFFFFFFFFFFFFFFFFFFFFFFFFFFFFFFFFFFFFFFFFFFFFFFFFFFFFFFFFFFFFFFFFFFFFFFFFFFFFFFFFFFFFFFFFFFFFF
{            \begin{figure}[ht]
\centering
\includegraphics[width=0.55\textwidth]{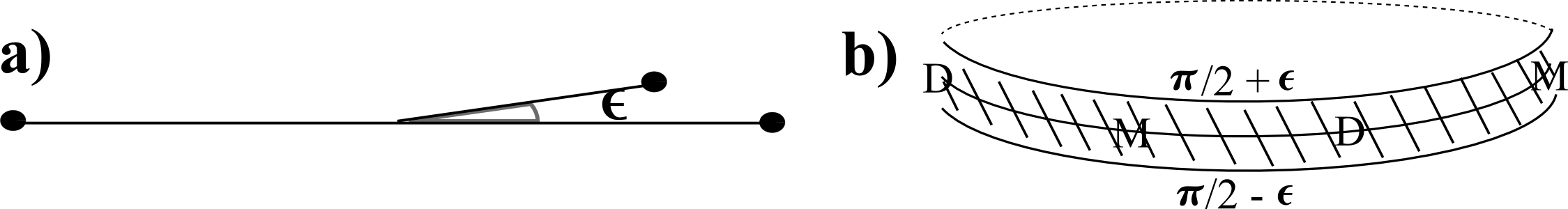}
\caption[Text der im Bilderverzeichnis auftaucht]{        \footnotesize{a) $\epsilon$-collinearity in 
space for three points.  b) The corresponding configuration space belt of width 2$\epsilon$.} }
\label{Fig7a} \end{figure}          }
%FFFFFFFFFFFFFFFFFFFFFFFFFFFFFFFFFFFFFFFFFFFFFFFFFFFFFFFFFFFFFFFFFFFFFFFFFFFFFFFFFFFFFFFFFFFFFFFFFFFFFFFF

\noindent Next, I consider the role of configuration space regions in a number of these approaches. 
I find that Gibbons--Pope type coordinates are useful in considering regions of configuration space: the volume 
element is simple in these, 

\noindent $\mbox{sin}^3\chi\mbox{cos}\,\chi\,\d\chi\mbox{sin}\,\beta\, \d\beta \, \d\phi \, \d\psi/8$, 
and so are characterizations for a number of physically and geometrically significant regions. 

\mbox{ } 

\noindent Application 1) To computing na\"{i}ve Schr\"{o}dinger interpretation probabilities of `being', via (\ref{NSI}).  
This is a continuation of what I have done in previous papers for metrolands and triangleland \cite{AF, +tri, ScaleQM, 08III}.  
Note the `sphere factor' within this volume element (the $\beta$ and $\phi$ factors).  
The above gives 
\beq
\mbox{Prob}(\mbox{Region R}) \propto \int_{\sR}|\Psi(\chi,\beta,\phi,\psi)|^2  
\mbox{sin}^3\chi\mbox{cos}\,\chi\,\d\chi\mbox{sin}\,\beta\,\d\beta\, \d\phi \, \d\psi \mbox{ } . 
\eeq
Then both volume element and the wavefunctions (at least in the free potential case \cite{MacFarlane, Quad1}) 
separate into a product of factors $C(\chi)D(\beta)E(\phi)H(\psi)$, 
and also the region of integration (at least for a number of physically-significant regions including the below examples)
so that the integration over configuration space reduces to a product of 1-$d$ integrals, 
rendering integration relatively straightforward.

The next issue addressed in this paper is to characterize some physically-significant R's.  
[The wavefunctions themselves are provided in \cite{Quad1}, where I combine them and this paper's study of 
regions to also provide na\"{\i}ve Schr\"{o}dinger interpretation probabilities for 
approximate-collinearity, approximate-squareness and approximate-triangularity.]

4-stop metroland and triangleland counterparts of such regions are e.g. caps, belts and lunes in spherical polar 
coordinates \cite{AF, +tri} endowed with particular physical significance, such as the cap of $\epsilon$-equilaterality 
or the belt of 

\noindent $\epsilon$-collinearity [Fig 8b)]. 
The present paper's regions are somewhat more involved due to involving the more complicated and higher-dimensional 
geometry of $\mathbb{CP}^2$.  
One thing noted for the shape space spheres is a correspondence between the size of the region in question 
(e.g. the radius of a small cap, the width of a small belt, the angle of a narrow lune and combinations of these 
by union, intersection and negation) and the size of the departure in space from the precise configuration.  
E.g. the width of the belt of collinearity on the shape sphere corresponds to a Kendall type \cite{Kendall, +tri} 
notion of $\epsilon$-collinearity of three points in space (Fig 8).  
Establishing such correspondences is then part of the study of physically-significant regions of the $\mathbb{CP}^2$ 
configuration space also.  
In particular, I consider the following.  

\mbox{ } 

\noindent
I) Approximately-collinear quadrilaterals.  
Exactly collinear was $\mathbb{RP}^2$ corresponding to, in configuration space studied using H-coordinates, 
both angular coordinates $\phi$ and $\psi$ being 0 or an integer multiple of $\pi$.  
This condition is now $\epsilon$-relaxed.  
Thus the region of integration is all values of the ratio coordinates $\beta$ and $\chi$ whilst the angular coordinates are to 
live within the following union of products of intervals: 
$$
\big(
\{0 \leq \phi \leq \epsilon\} \cup \{\pi - \epsilon \leq \phi \leq \pi + \epsilon\} \cup \{2\pi - \epsilon \leq \phi \leq 2\pi\}
\big) 
\times
$$
\beq
\big(
\{0 \leq \psi \leq \epsilon\} \cup \{\pi - \epsilon \leq \psi \leq \pi + \epsilon\} \cup \{2\pi - \epsilon \leq \phi \leq 2\pi + \epsilon\}
\cup 
\{3\pi - \epsilon \leq \phi \leq 3\pi + \epsilon\} \cup \{4\pi - \epsilon \leq \phi \leq 4\pi\}
\big) \mbox{ } .  
\eeq
From the perspective of each configuration in space, this notion of $\epsilon$-collinearity corresponds to the relative angles 
$\Phi_1$ and $\Phi_2$ each lying within $[0, \epsilon] \cup [\pi - \epsilon, \pi + \epsilon] \cup [2\pi - \epsilon, 2\pi]$.

\mbox{ }

\noindent II) Quadrilaterals that are approximately triangular or approximately one of the mergers depicted in the 
coarse-graining triangles or parallelogram of Fig 7.     
The exact configuration in each of these cases is a sphere characterized by one ratio variable and one relative angle variable.  
In moving away from exact triangularity, a second $\epsilon$-sized ratio variable becomes involved, and, 
in doing so one is rendering the other relative angle meaningful, allowing it to take all values.  Thus the region 
of integration here is all angles, one ratio variable taking all possible values also, and the other being confined 
to an $\epsilon$-interval about the value that corresponds to exact triangularity.    
To be more concrete, consider the \{+43\} triangle in Gibbons--Pope type coordinates that derive from H-coordinates.  
Here, $\beta = \pi$, so the region of integration is all $\phi$, all $\psi$, all $\chi$ and $\pi - \epsilon \leq \beta \leq \pi$.
Then, for example, the notion of $\epsilon$-\{+43\} triangular corresponds in space to the ratio \{12\}/\{+$\times$\} 
being of size $\epsilon/2$ or less.  

\mbox{ }

\noindent 
III) One of the labellings of exact square configuaration is at $\phi = 0$, $\psi = \pi$, $\beta = \pi/2$ and $\chi = \pi/4$.
Approximate squareness allows for all four of these quantities to be $\epsilon$-close, so that the region of integration is a 
product of four intervals of width $\epsilon$ about these points.  
That corresponds to the two sides of the Jacobi H being allowed to be $\epsilon$-close ($\beta$-variation), the quadrilateral to 
vary in height to length ratio ($\chi$-variation) and for each of the sides of the Jacobi H to become non-right with respect to 
the cross-bar ($\psi$- and $\phi$-variation), which is an entirely intuitive parametrization of the possible small 
departures from exact squareness.
[If one is interested in all of the labellings of the square, one can straightforwardly characterize each with a similar 
construction and again take the union of these regions.]  
This example is clearly furthermore useful as a notion of approximate uniformity, which is of interest in classical and 
quantum Cosmology.  

\mbox{ }

\noindent 
Finally, note that if one considers mirror-image-identified and/or indistinguishable particles, then one has just a portion of 
$\mathbb{CP}^2$ in total and then the physically significant regions are smaller subsets (including sums over less things, 
e.g. there are now 3, 2 or just 1 distinct square configurations).  

\mbox{ } 

\noindent Application 2) The above notions of closeness and use of the Fubini--Study kinetic metric additionally embody 
control over localization both in space and in configuration space, allowing for the triangleland work towards establishing 
a records theory in \cite{FileR} (concerning in particular notions of distance between shapes) to be extended to quadrilateralland.  

\mbox{ }

\noindent Application 3) In the semiclassical approach, the many approximations used only hold in certain regions, in 
particular the crucial Born--Oppenheimer and WKB approximations.  
The semiclassical application awaits having the semiclassical approach sorted out for scaled quadrilateralland 
\cite{Quad1}; for the moment see \cite{SemiclIII} for the triangleland counterpart of such workings.  
This is needed for extending semiclassical work in \cite{ScaleQM, 08III, SemiclIII} to the quadrilateralland setting.  

\mbox{ } 

\noindent Application 4) Regions also feature in the histories approach and in Halliwell's combination of this 
with semiclassical approach and timeless ideas. 
Indeed, an extra connector between these approaches is that \cite{H03, H09} the semiclassical approach aids in the 
computation of timeless probabilities of histories entering given configuration space regions.
This, by the WKB assumption, gives a wavefunction flux into each region in terms of $\fW$ and the Wigner function (see e.g. \cite{H03}). 
Moreover, such schemes go beyond the standard semiclassical approach, so there is some chance that
further objections to the semiclassical approach (problems inherited from the Wheeler--DeWitt equation and problems with 
reconstructing spacetime in such approaches) would be absent from the new unified strategy.
I am presently considering extending \cite{H03} to RPM's (for the moment for triangleland), looking at, without reference to time, 
what is the probability  of finding the system in a series of regions of configuration space for a given eigenstate of the Hamiltonian \cite{H03}?
Halliwell studied this with a free particle, a working which has a direct, and yet more genuinely closed-universe, 
counterpart for scaled triangleland \cite{08I} via the `Cartesian to Dragt coordinates correspondence'  
allowing me to transcribe this working to a relational context.
The quadrilateralland extension of this calculation, whilst remaining relationally interpretable 
as a closed-universe model, would represent a step-up in complexity and mathematical novelty for this program.

\mbox{ }

\noindent {\bf Acknowledgements}: I thank those close to me for being supportive of me whilst this work was done.   
Professors Don Page and Gary Gibbons for teaching me about $\mathbb{CP}^2$.
Dr Julian Barbour for introducing me to RPM큦.  
Professors Gary Gibbons, Jonathan Halliwell, Chris Isham and Karel Kucha\v{r} for discussions. 
Mr Eduardo Serna for reading earlier drafts of this manuscript.  
Professors Belen Gavela, Marc Lachi\`{e}ze-Rey, Malcolm MacCallum, Don Page, Reza 
Tavakol, and Dr Jeremy Butterfield for support with my career.  
Fqxi grant RFP2-08-05 for travel money whilst part of this work was done in 2009-2010, 
and Universidad Autonoma de Madrid for funding in 2010--2011.

%=====================================================BIBLIOGRAPHY==========================================================================


\begin{thebibliography}{99}
%===========================================================================================================================================

\footnotesize

\bibitem{BB82}               J.B. Barbour and B. Bertotti, Proc. Roy. Soc. Lond. {\bf A382} 295 (1982). 

\bibitem{B03}                J.B. Barbour, Class. Quantum Grav. \textbf{20} 1543 (2003), gr-qc/0211021. 

\bibitem{B94I}               J.B. Barbour, Class. Quantum Grav. {\bf 11} 2853 (1994).

\bibitem{EOT}                J.B. Barbour, {\it The End of Time} (Oxford University Press, Oxford 1999).

\bibitem{Rovellibook}        C. Rovelli, {\it Quantum Gravity} (Cambridge University Press, Cambridge 2004).    

\bibitem{08I}                E. Anderson, Class. Quantum Grav. {\bf 26} 135020 (2009), arXiv:0809.1168.   

\bibitem{RWR}                J.B. Barbour, B.Z. Foster and N. \'{O} Murchadha, Class. Quantum Grav. 
                             {\bf 19} 3217 (2002), gr-qc/0012089; 
%
                             E. Anderson, Gen. Rel. Grav. {\bf 36} 255, gr-qc/0205118; 
% 
                             Phys. Rev. {\bf D68} 104001 (2003),  gr-qc/0302035;
%
                             ``Geometrodynamics: Spacetime or Space?"  
                             (Ph.D. Thesis, University of London 2004), gr-qc/0409123;
%
                             E. Anderson, in {\it General Relativity Research Trends, Horizons in World 
                             Physics} {\bf 249} ed. A. Reimer (Nova, New York 2005), gr-qc/0405022;
%
                             Stud. Hist. Phil. Mod. Phys. {\bf 38} 15 (2007), gr-qc/0511070;
%
                             in ``Classical and Quantum Gravity Research", ed. M.N. 
                             Christiansen and T.K. Rasmussen (Nova, New York 2008), arXiv:0711.0285.  

\bibitem{FORD}               E. Anderson, Class. Quantum Grav. {\bf 25} 025003 (2008), arXiv:0706.3934.

\bibitem{Lanczos}            C. Lanczos, {\it The Variational Principles of Mechanics} (University of 
                             Toronto Press, Toronto 1949). 

\bibitem{BSW}                R.F. Baierlain, D. Sharp and J.A. Wheeler, Phys. Rev. {\bf 126} 1864 (1962).

\bibitem{FEPI}               E. Anderson, Class. Quantum Grav. {\bf 25} 175011 (2008), arXiv:0711.0288.

\bibitem{Cones}              E. Anderson, arXiv:1001.1112.

\bibitem{Kendall}            D.G. Kendall, D. Barden, T.K. Carne and H. Le, {\it Shape and Shape Theory} (Wiley, Chichester 1999).  

\bibitem{K92}                K.V. Kucha\v{r}, in {\it Proceedings of the 4th Canadian Conference on 
                             General Relativity and Relativistic Astrophysics} ed. G. Kunstatter, D. 
                             Vincent and J. Williams (World Scientific, Singapore 1992). 

\bibitem{Kieferbook}         
                             See e.g. C. Kiefer, {\it Quantum Gravity} (Clarendon, Oxford 2004).  

\bibitem{06I06IISemiclIGrybBanal}
                             E. Anderson, Class. Quantum Grav. {\bf 23} (2006) 2469, gr-qc/0511068; 
%
                             2491 (2006), gr-qc/0511069;
%
                             {\bf 24} 2935 (2007), gr-qc/0611007;
%
                             {\bf 27} 045002 (2010), arXiv:0905.3357;
%
                             S.B. Gryb, arXiv:0804.2900; 
%
                             Class. Quantum Grav. {\bf 26} (2009) 085015, arXiv:0810.4152;  
%
                             J.B. Barbour and B.Z. Foster, arXiv:0808.1223.  

\bibitem{ScaleQM}            E. Anderson, Class. Quantum Grav. {\bf 28} 065011 (2011), arXiv:1003.1973.

\bibitem{MGM}                E. Anderson, for Proceedings of Paris 2009 Marcel Grossman Meeting, in Press, arXiv:0908.1983.

\bibitem{Records}            E. Anderson, Int. J. Mod. Phys. {\bf D18} 635 (2009), arXiv:0709.1892;   
%
                             in {\it Proceedings of the Second Conference on Time and 
                             Matter}, ed. M. O'Loughlin, S. Stani\v{c} and D. Veberi\v{c} 
                             (University of Nova Gorica Press, Nova Gorica, Slovenia 2008), arXiv:0711.3174.   

\bibitem{AF}                 E. Anderson and A. Franzen, Class. Quantum Grav. {\bf 27} 045009 (2010), 
                             arXiv:0909.2436. 

\bibitem{Mini}               M. Ryan, {\it Hamiltonian Cosmology} (Lecture Notes in Physics 13) (Springer, Berlin, 1972); 
%
                             J.B. Hartle and S.W. Hawking, Phys. Rev. {\bf D28} 2960 (1983); 
% 
                             D.L. Wiltshire, in {\it Cosmology: the Physics of the Universe} 
                             ed. B. Robson, N. Visvanathan and W.S. Woolcock 
                             (World Scientific, Singapore 1996), gr-qc/0101003.   

\bibitem{Carlip}             S. Carlip, {\it Quantum Gravity in 2 + 1 Dimensions} 
                             (Cambridge University Press, Cambridge 1998).  

\bibitem{I93}                C.J. Isham, in {\it Integrable Systems, Quantum Groups and Quantum Field Theories}  
                             ed. L.A. Ibort and M.A. Rodr\'{\i}guez (Kluwer, Dordrecht 1993), gr-qc/9210011.

\bibitem{APOT}               E. Anderson, arXiv:1009.2157. 

\bibitem{HallHaw}            J.J. Halliwell and S.W. Hawking, Phys. Rev. {\bf D31}, 1777 (1985). 

\bibitem{08II}               E. Anderson, Class. Quantum Grav. {\bf 26} 135021 (2009), gr-qc/0809.3523.

\bibitem{+tri}               E. Anderson, Gen. Rel. Grav. {\bf 43} 1529 (2011), arXiv:0909.2439.  

\bibitem{08III}              E. Anderson, arXiv:1005.2507. 

\bibitem{QShape}             E. Anderson, arXiv:1009.2161 (Seminar I on relational quadrilaterals).   

\bibitem{Dragt}              A.J. Dragt, J. Math. Phys. {\bf 6} 533 (1965).

\bibitem{ZickACG86LR}        W. Zickendraht, Phys. Rev. {\bf 159} 1448 (1967); 
%
                             J. Math. Phys. {\bf 10} 30 (1969);                 
%
                             {\bf 12} 1663 (1970);
%
                             V. Aquilanti, S. Cavalli and G. Grossi, J. Chem. Phys. {\bf 85} 1362 (1986); 
%
                             R.G. Littlejohn and M. Reinsch, Rev. Mod. Phys. {\bf 69} 213 (1997).

\bibitem{LR95}               R.G. Littlejohn and M. Reinsch, , Phys. Rev. {\bf A52} 2035 (1995).

\bibitem{QCons}              E. Anderson, arXiv:1202.4187 (Seminar III on relational quadrilaterals).  

\bibitem{Quad1}              E. Anderson and S.A.R. Kneller, arXiv:1303.5645.

\bibitem{Kuiper}             N.H. Kuiper, Math. Ann. {\bf 208} 175 (1974).  

\bibitem{GiPo}               G.W. Gibbons and C.N. Pope, Commun. Math. Phys. {\bf 61} 239 (1978); 
%
                             C.N. Pope, Phys. Lett. {\bf 97B}  417 (1980).  
 
%%%%%%%%%%%%%%%%%%%%%%%%%%%%%%%%%%%%%%  B O D Y %%%%%%%%%%%%%%%%%%%%%%%%%%%%%%%%%%%%%%%%%%%%%%%%%%%%%%%% 

\bibitem{Marchal}            See e.g. C. Marchal, {\it Celestial Mechanics} (Elsevier, Tokyo 1990).

\bibitem{Battelle}            J.A. Wheeler, in {\it Battelle Rencontres: 1967 Lectures in Mathematics 
                              and Physics} ed. C. DeWitt and J.A. Wheeler (Benjamin, New York 1968).

\bibitem{York73York74ABFOABFKO} 
                             J.W. York, Phys. Rev. Lett. {\bf 28} 1082 (1972); 
%
                             J. Math. Phys. {\bf 14} 456 (1973);
%
                             Ann. Inst. Henri Poincar\'{e} {\bf 21} 319 (1974); 
%
                             E. Anderson, J.B. Barbour, B.Z. Foster and N. \'{O} Murchadha, Class. Quantum Grav. {\bf 20} 157 (2003), gr-qc/0211022.  
%
                             E. Anderson, J.B. Barbour, B.Z. Foster, B. Kelleher and N. \'{O} Murchadha, 
                             Class. Quantum Grav {\bf 22} 1795 (2005), gr-qc/0407104.   

\bibitem{NewBO}              J. Barbour and N. \'{O} Murchadha, arXiv:1009.3559.  

\bibitem{FM96}               A.E. Fischer and V. Moncrief, Gen. Rel. Grav.  {\bf 28}, 221 (1996).

\bibitem{FileR}               E. Anderson, arXiv:1111.1472.  

\bibitem{WCP}             The following consider a weighted projective space $\mathbb{WCP}^3 = \mathbb{CP}^3/\mathbb{Z}_2$ in 
the context of M-theory. 
%
E. Witten, hep-th/0108165; 
%
M. Atiyah and E. Witten, hep-th/0107177;  
%
B.S. Acharya and E. Witten, hep-th/0109152; 
%
B.S. Acharya, in {\it Strings and Geometry Proceedings of the Clay 
                             Mathematics Institute 2002 Summer School)} ed. M. Douglas, J. Gauntlett 
                             and M. Gross (American Mathematical Society, Providence, Rhode Island 2003), 
                             available online at 
                             http://www.claymath.org/library/proceedings/cmip03c.pdf;   
%
D. Joyce, ibid,  mathDG/9910002; 
%
A. Collinucci, JHEP 0908:076 (2009), arXiv:0812.0175. mentions a $\mathbb{WCP}^4$;
%
R. Auzzi, M. Shifman and A. Yung Phys. Rev. {\bf D73} 105012 (2006); 
% 
                             Erratum-ibid. {\bf D76} 109901 (2007), hep-th/0511150 also consider $\mathbb{WCP}^5$;
% 
E. Witten, Adv. Theor. Math. Phys. {\bf 5} 841 (2002)  hep-th/0006010 considers a $\mathbb{WCP}^k$ and  
%
 M. Eto, K. Konishi, G. Marmorini, M. Nitta, K. Ohashi, W. Vinci and  
                             N. Yokoi, Phys. Rev. {\bf D74} 065021 (2006), hep-th/0607070 state that some such spaces are singular. 

\bibitem{Veronese}           See e.g. J. Harris, {\it Algebraic Geometry, A First Course,} (Springer-Verlag, New York 1992).

\bibitem{Isham84}            C.J. Isham, in {\it Relativity, Groups and Topology {II}} ed. B.S. DeWitt 
                             and R. Stora (North-Holland, Amsterdam 1984). 

\bibitem{Riemann}            See e.g. A.F. Beardon, {\it Primer on Riemann Surfaces} (Cambridge University Press, Cambridge 1984).  

%%%%%%%%%%%%%%%%%%%%%  C O N C L U S I O N   R E F E R E N C E S %%%%%%%%%%%%%%%%%%%%%%%%%%%%%%%%%%%%%%%%

\bibitem{Semi}               B.S. DeWitt, Phys. Rev. {\bf 160} 1113 (1967);
%
                             V.G. Lapchinski and V.A. Rubakov, Acta Physica Polonica {\bf B10} (1979);
%
                             T. Banks, Nu. Phys. {\bf B249} 322 (1985).  

\bibitem{SemiclIII}          E. Anderson, arXiv:1101.4916.

\bibitem{Forth}              E. Anderson, arXiv:1305.4685; 
%
``Quantum Cosmology will become a Numerical Subject", Invited Seminar at 'XXIX-th International Workshop on High Energy Physics: 
							 New Results and Actual Problems in Particle \& Astroparticle Physics and Cosmology', Moscow 2013, arXiv:1306.5812.  

\bibitem{HP86UW89}           S.W. Hawking and D.N. Page,  Nu. Phys. {\bf B264} 185 (1986); 
%
                             W. Unruh and R.M. Wald, Phys. Rev. {\bf D40} 2598 (1989). 

\bibitem{PW83}                D.N. Page and W.K. Wootters, Phys. Rev. {\bf D27}, 2885 (1983).                 

\bibitem{GMH}                 M. Gell-Mann and J.B. Hartle, Phys. Rev. {\bf D47} 3345 (1993).       

\bibitem{H99}                 J.J. Halliwell, Phys. Rev. {\bf D60} 105031 (1999), quant-ph/9902008.

\bibitem{Hartle}              J.B. Hartle, in {\it Gravitation and Quantizations: 
                              Proceedings of the 1992 Les Houches Summer School} ed. B. Julia and 
                              J. Zinn-Justin (North Holland, Amsterdam 1995),  gr-qc/9304006. 

\bibitem{H03}                J.J. Halliwell, in {\it The Future of Theoretical Physics and Cosmology  
                             (Stephen Hawking 60th Birthday Festschrift Volume)} ed. G.W. Gibbons, 
                             E.P.S. Shellard and S.J. Rankin (Cambridge University Press, Cambridge 2003), 
                             gr-qc/0208018.  

\bibitem{H09}                 J.J. Halliwell, Phys. Rev. {\bf D80} 124032 (2009), arXiv:0909.2597.

\bibitem{MacFarlane}         A.J. MacFarlane, J. Phys. A: Math. Gen. {\bf 36} 7049 (2003).

\end{thebibliography}
\end{document}